\documentclass[aip,graphicx]{revtex4-1}
\usepackage{graphicx}
\usepackage{amsfonts}
\usepackage{amsmath}
\usepackage{dsfont}

\draft 

\begin{document}


\title[]{$\pi$-Girsanov: A Generalized Method to Construct Markov State Models from Non-Equilibrium and Multiensemble Biased Simulations}

\author{Mingyuan Zhang}
\affiliation{ 
Institut de Qu\'{\i}mica Teor{\`i}ca i Computacional (IQTCUB), Universitat de Barcelona, Barcelona 08028, Spain
}%

\author{Yong Wang}
\affiliation{ 
College of Life Sciences, Zhejiang University, Hangzhou 310027, China}

\author{Bettina G.~Keller}%
\email{bettina.keller@fu-berlin.de}
\affiliation{%
Freie Universität Berlin, Department of Biology, Chemistry and Pharmacy, Arnimallee 22, 14195 Berlin
}%

\author{Hao Wu}
\email{hwu81@sjtu.edu.cn}
\affiliation{%
School of Mathematical Sciences, Institute of Natural Sciences, and MOE-LSC, Shanghai Jiao Tong University, Shanghai, China
}%

\begin{abstract}
We introduce $\pi$-Girsanov, a new method for constructing Markov state models from biased enhanced-sampling molecular dynamics simulations based on Girsanov reweighting. The key idea behind this new method is to separate the reweighting of the stationary density from the reweighting of the correlation function. We evaluate the effectiveness of this approach on several analytical potentials and on a model biomolecular system, comparing its performance with the original method. Our results show that $\pi$-Girsanov not only improves the estimation in a single-ensemble setting, but also resolves key challenges in estimating transition matrices from multiensemble and non-equilibrium biased trajectories. Overall, $\pi$-Girsanov represents a substantial advance in kinetic reweighting, strengthening the connection between enhanced sampling techniques and Markov state modeling.
\end{abstract}


\maketitle 

\section{Introduction}
\label{sec:intro}
Markov state models (MSM) \cite{prinz2011markov, chodera2007automatic, buchete2008coarse} have now become a routine and powerful tool for analyzing complex high-dimensional molecular dynamics (MD) simulation trajectories. These models allow us to access the thermodynamic and kinetic properties and identify the major metastable states and transition pathways to obtain valuable physical insights of the system. MSMs have proven useful for studying a wide range of biomolecular systems, including protein folding\cite{schwantes2013improvements,mckiernan2017modeling,paul2021thermodynamics,nagel2023toward,zhang2024flow}, protein-protein interaction\cite{schor2015shedding,plattner2017complete}, ligand dissociation \cite{plattner2015protein,paul2017protein}. 

The common procedures for building an MSM from unbiased MD trajectories are as follows \cite{prinz2011markov}. First, one discretizes the entire phase space into individual non-overlapping microstates via a Voronoi tessellation. Then, one computes the cross correlation matrix $\mathbf C(\tau)$, or equivalently the count matrix, which corresponds to the total number of transitions between all these microstates at a chosen lag time $\tau$ observed in the MD trajectories. Finally, given the observation $\mathbf C(\tau)$, one estimates a transition matrix $\mathbf T(\tau)$, which contains the transition probabilities between all microstates, with a maximum likelihood \cite{bowman2009progress,prinz2011markov,trendelkamp2015estimation} or Bayesian approach\cite{trendelkamp2013efficient,trendelkamp2015estimation}.

One notable issue of this protocol is that the convergence of the estimation of $\mathbf C(\tau)$ is, in general, very slow. This is due to systems propagated by MD simulations often remaining trapped in a few metastable states separated by high-energy barriers and generating highly correlated samples for the estimation of $\mathbf C(\tau)$. As a result, brute force unbiased MD sampling using currently available hardware obtains $\mathbf C(\tau)$ with high variance, which then contaminates the downstream evaluations of the transition matrix, thermodynamic and kinetic properties of the system.

A general and well-established method of variance reduction for MD simulation is to employ enhanced sampling \cite{henin2022enhanced}. Instead of propagating the system at the original potential, one can simulate the system with the addition of a bias potential. The bias potential is often constructed in a way that lowers the energy barriers and promotes the transitions between system's metastable states to improve the statistical efficiency of MD simulation. Of course, the introduction of a bias potential will inevitably alter the system's kinetic and thermodynamic properties. Yet, for thermodynamic properties like the equilibrium Boltzmann distribution, which are phase space ensemble averages, multiple statistical reweighting techniques exist, which assign a weight to each sample generated by the biased simulations to correct this bias effect. Examples are weighted histogram analysis method (WHAM) \cite{kumar1992weighted,souaille2001extension}, multistate Bennett acceptance ratio (MBAR) \cite{shirts2008statistically}, dynamic histogram analysis method (DHAM) \cite{rosta2015free}, transition-based reweighting analysis method (TRAM) \cite{wu2016multiensemble}. Over the past decades, many enhanced sampling algorithms for MD simulations have been developed following such ideas, such as umbrella sampling \cite{torrie1977nonphysical}, steered MD\cite{park2003free}, metadynamics\cite{laio2002escaping,barducci2008well,bussi2020using}, and they have been successfully applied to compute the thermodynamic properties of various biomolecular systems of interest. 

From this perspective, a natural idea to reduce the variance of $\mathbf C(\tau)$ is to simulate the system similarly with bias potential and employ statistical reweighting to estimate the unbiased $\mathbf C(\tau)$. However, the elements of $\mathbf C(\tau)$, $c_{ij}(\tau)$, are cross correlation functions that compute the time correlation between two different observables separated by a lag time $\tau$ and thus are kinetic properties instead of thermodynamic properties. The corresponding statistical reweighting procedures for kinetic properties from biased simulations are still an active area of research \cite{keller2024dynamical} and a widely-used, reliable approach has not been established.

One promising attempt in tackling this challenge is the Girsanov reweighting technique \cite{donati2017girsanov,donati2018girsanov,kieninger2023girsanov} based on Girsanov's theorem \cite{girsanov1960transforming}, 
which specifies how statistical weights can be assigned to stochastic trajectories under changes of the underlying dynamics.
Kinetic properties like the cross correlation functions can be interpreted as a combined ensemble average over both the phase space and the path space (see Section \ref{sec:theory}). The phase space ensemble average, as stated above, can be reweighted using established methods in principle. The reweighting of the path ensemble average, on the other hand, requires the analytical solution of a path probability ratio, which has been provided for both overdamped \cite{zuckerman2000efficient, athenes2004path, xing2006calculation, donati2017girsanov} and underdamped \cite{kieninger2023girsanov} Langevin dynamics. Following such theoretical progress, Girsanov reweighting has been successfully combined with the MSM methodology\cite{schutte2015markov, donati2017girsanov} and applied to various analytical and molecular systems \cite{donati2017girsanov,donati2018girsanov,schafer2024implementation, shmilovich2023girsanov, bolhuis2025optimal}. Furthermore, the Girsanov reweighting algorithm has also been implemented in the MD simulation code OpenMM and the MSM code Deeptime to improve their accessibility to a broader audience\cite{schafer2024implementation}.

Nevertheless, room for improvement exists for the current Girsanov reweighted MSM method. For instance, previous studies\cite{donati2017girsanov,donati2018girsanov,schafer2024implementation} have limited their applications to mainly trajectories simulated in a single biased ensemble with a static bias potential due to the difficulties in correctly reweighting multiensemble biased simulations or non-equilibrium simulations with a time-dependent bias potential. However, popular enhanced sampling algorithms mentioned above belong to either of these categories: umbrella sampling produces multiensemble simulations, whereas metadynamics and steered MD produce non-equilibrium simulations. It is therefore of great significance to develop a more general version of Girsanov reweighted MSM method which is able to cope with biased trajectories from these simulations.

In this paper, we present a novel method to construct MSMs with biased trajectories, which we refer to as the $\pi$-Girsanov method. In $\pi$-Girsanov, we combine (i) thermodynamic reweighting of the unbiased Boltzmann distribution, (ii) Girsanov-based reweighting of path-ensemble averages, and (iii) a stationary-vector-constrained maximum-likelihood estimator for the transition matrix developed in Ref.\citenum{trendelkamp2015estimation}. This combination enables a generalized computation of transition matrices from biased trajectories generated from multiple simulations with different time-dependent or time-independent biases. We demonstrate on analytical potentials that the proposed method is capable of reweighting biased trajectories correctly from various popular enhanced sampling algorithms for the construction of MSMs, even though the performance of using trajectories from certain enhanced sampling simulations is significantly better. We also demonstrate that even for a single ensemble biased simulation with a static bias potential, the current methodology is able to generate a better estimation of the transition matrix with less variance than the previous method. We then applied our method to the trajectories of simulating a model biomolecular system under multiple different time-dependent biases and successfully obtained the correct thermodynamics and kinetics of the system. Finally, we further discuss the limitations of the current protocol and how this can be potentially improved. Our contribution is a major step in the development of statistical reweighting techniques for the construction of MSMs with enhanced sampling trajectories and their subsequent applications in understanding important biomolecular systems.

\section{Theory}
\label{sec:theory}
\subsection{Molecular Dynamics}
We restrict our discussion in this paper to MD simulations in the canonical ensemble propagated by underdamped Langevin dynamics. Such setting is widely applicable to the simulation of solvated biomolecular systems and provides more accurate modeling of molecular kinetics in comparison to overdamped Langevin dynamics. For an $N$-particle molecular system, the time evolution of the phase space coordinate $x = (p,q) \in  \Gamma \subset \mathbb R^{6N}$ modeled by underdamped Langevin dynamics under a potential $U(q)$ can be written as:
\begin{equation}
\begin{aligned}
\mathrm dq_t &= M^{-1}p_t \mathrm dt\\
\mathrm dp_t &= -\nabla U(q_t) \mathrm dt - \xi p_t\mathrm dt + \sqrt{2k_BT\xi M} \mathrm dW_t
\end{aligned}
\label{eq:1}
\end{equation}
where we use $q_t \in \mathbb R^{3N}$ to denote the Cartesian position coordinate of the system at time $t$, $p_t \in \mathbb R^{3N}$ to denote the momentum of the system at time $t$ and $\Gamma$ to denote the phase space. Furthermore, $M$ is the $3N\times 3N$ diagonal mass matrix,  $\xi$ is the collision rate, $k_B$ is the Boltzmann constant, $T$ is the temperature and $\mathrm d W_t$ is a $3N$-dimensional standard Wiener process. The system propagated by (\ref{eq:1}) has a unique stationary probability density $\pi(x)$, known as the Boltzmann distribution:
\begin{equation}
\pi(x) \propto e^{-\frac{\mathcal H(x)}{k_BT}}
\label{eq:2}
\end{equation}
where $\mathcal{H}(x) = K(p) + U(q)$ is the system Hamiltonian, $K(p) = \frac{1}{2}p^{\top}M^{-1}p$ is the kinetic energy and $U(q)$ is the potential energy function.

For general molecular systems, (\ref{eq:1}) can only be solved by numerical integration. For example, for the ABOBA integrator \cite{leimkuhler2015molecular} we used throughout this paper, the update cycle of the momentum $p_{t+\Delta t}$ and coordinate $q_{t+\Delta t}$ from $p_t$ and $q_t$ after a small time step $\Delta t$ is split into 5 individual steps:
\begin{equation}
\begin{aligned}
q_{t+\frac{1}{2}\Delta t} &= q_t + \frac{1}{2}\Delta tM^{-1}p_t\\
p_{t+\frac{1}{2}\Delta t} &= p_t - \frac{1}{2}\Delta t\nabla U(q_{t+\frac{1}{2}\Delta t})\\
\hat p_{t+\frac{1}{2}\Delta t} &= e^{-\xi\Delta t}p_{t+\frac{1}{2}\Delta t}+\sqrt{k_BTM(1-e^{-2\xi\Delta t})}\cdot \eta_t\\
p_{t+\Delta t} &= \hat p_{t+\frac{1}{2}\Delta t} - \frac{1}{2}\Delta t\nabla U(q_{t+\frac{1}{2}\Delta t})\\
q_{t+\Delta t} &= q_{t+\frac{1}{2}\Delta t} + \frac{1}{2}\Delta t M^{-1}p_{t+\Delta t}
\end{aligned}
\label{eq:3}
\end{equation}
Here, $\eta_t\sim \mathcal N(0,I)\in \mathbb R^{3N}$ is a standard Gaussian random variable with zero mean and unit variance, sampled independently at each $t$. One thing worth noticing in (\ref{eq:3}) is that the integration of the ABOBA integrator over time is a random Markovian process. The randomness of this process arises only from the sampling of $\eta_t$ in the third step of each integration cycle. Therefore, we can see that the conditional probability $\rho(x_{t+\Delta t}|x_t)$ of observing $x_{t+\Delta t} = (p_{t+\Delta t},q_{t+\Delta t})$ at $t+\Delta t$, given we have $x_t = (p_t,q_t)$ at $t$ can be related to the probability $\rho(\eta_t)$ of sampling a specific $\eta_t$ from $\mathcal N(0,I)$, which is analytically available (see Appendix for details):
\begin{equation}
\rho(x_{t+\Delta t}|x_t) \propto \rho(\eta_t) = \prod^{3N}_{i=1}\frac{1}{\sqrt{2\pi}}e^{-\frac{\eta_{t,i}^2}{2}}
\label{eq:4}
\end{equation}
where $\eta_{t,i}$ is the $i$th element of $\eta_t$. Such relationship in (\ref{eq:4}) is the foundation of Girsanov reweighting, which we will elaborate below.

\subsection{Markov State Modeling}
To build an MSM from the unbiased trajectories generated by (\ref{eq:3}) under $U(q)$, we can first consider these trajectories as a collection of paths $\omega$. A path $\omega = (x_1,x_2,...,x_\tau|x_0)$ is defined as a realization of the dynamics in (1) with a given starting point $x_0$ in the phase space $\Gamma$. The associated path space of $\omega$ is denoted as $\Omega \subset \mathbb R^{6N\cdot \tau}$. 

Then, one discretize the whole phase space $\Gamma$ into $L$ disjoint microstates $S = (S_1,S_2,...,S_L) \subset \Gamma$ so that $S_i \cap S_j = \emptyset$ for all $i$ and $j$. As demonstrated in Ref.\citenum{prinz2011markov}, such discretization should ensure kinetically distant configurations are assigned to distinct microstates to avoid discretization error. Poorly discretized phase space will result in a systematic underestimate of the eigenvalues of the MSM and the timescales of the system slow modes\cite{noe2013variational}. For high-dimensional systems, the well-established protocol in practice is to first perform dimensionality reduction, either based on prior knowledge of the system or with data-driven techniques\cite{perez2013identification,noe2015kinetic,zhang2024flow}, so that the distance between configurations in the reduced space is approximately proportional to their kinetic distance\cite{noe2015kinetic}. State discretization is then performed in the reduced space using a grid, or unsupervised clustering algorithms such as K-means to define the microstates. Notice the discretization error of MSM can also be systematically improved by replacing such Voronoi tessellations with core-set discretizations \cite{schutte2011markov, lemke2016density} or more general ansatz functions \cite{nuske2014variational, schwantes2015modeling, mardt2018vampnets}. We here confine the discussion to MSMs based on non-overlapping microstates defined with the procedures described above.

Given the discretization, the cross-correlation matrix $\mathbf C(\tau)$ can then be computed from all path samples $\omega$ at a specific lag time $\tau$, with each of its entry $c_{ij}$ as a combined ensemble average over both the phase space and the path space\cite{donati2017girsanov}:
\begin{equation}
c_{ij}
= \int_\Gamma\pi(x_0)\mathds{1}_{S_i}(x_{0})\int_\Omega \mu(\omega)\mathds{1}_{S_j}(x_{\tau})\mathrm d\omega\mathrm dx_0= \lim_{m\to\infty}\left[\frac{1}{m}\sum_{k=1}^{m}\left(\mathds{1}_{S_i}(x_{0}^{(k)})\mathds{1}_{S_j}(x_{\tau}^{(k)})\right)\right]
\label{eq:5}
\end{equation}
Here, $\mu(\omega)$ is the conditional probability density of observing a path $\omega$ given its starting point at $x_0$, $m$ is the total number of paths collected, $x_0^{(k)}$, $x_\tau^{(k)}$ are the starting and end points of the $k$th path sample, $\mathds{1}_{S_i}(\cdot)$,$\mathds{1}_{S_j}(\cdot)$ are indicator functions of whether the phase space coordinates $x_0$ or $x_\tau$ belong to the $i$th or $j$th state and $\int_\Omega\mathrm d\omega$,$\int_\Gamma\mathrm dx_0$ are integrals of all $\omega$ or $x_0$ over the whole path space $\Omega$ or phase space $\Gamma$, respectively. Notice (\ref{eq:5}) does not imply that we have to run independent simulations starting from a collection of $x_0^{(k)}$. The path samples $\omega$ can be obtained from one or multiple regular long MD trajectories with a ``sliding window" scheme, as documented in Ref.\citenum{prinz2011markov}.

Given the computed $\mathbf C(\tau)$, one can then estimate its corresponding transition matrix $\mathbf T(\tau)$, with each of its element $\rho_{ij}$ defined as the conditional transition probability of observing a transition to the microstate $S_j$ after a lag time $\tau$, given the system is in the microstate $S_i$ at time $t$:
\begin{equation}
\rho_{ij} = \rho(x_{t+\tau} \in S_j | x_t \in S_i)
\label{eq:6}
\end{equation}
This is a statistical inference problem and there are multiple algorithms for solving $\mathbf T(\tau)$, based on either maximum likelihood estimation (MLE) \cite{anderson1957statistical,buchete2008coarse,bowman2009progress,prinz2011markov,trendelkamp2015estimation} or maximum a posteriori (MAP)\cite{trendelkamp2013efficient,trendelkamp2015estimation}. In this paper, we solely consider the estimation of MSMs with MLE. MLE suggests one can solve for the optimal set of parameters $\{\rho_{ij}^*\}$ of $\mathbf T(\tau)$ by maximizing the log-likelihood of observing $\mathbf C(\tau)$ for a given set of parameters $\{\rho_{ij}\}$. In mathematical terms, this can be expressed as \cite{anderson1957statistical,buchete2008coarse,prinz2011markov}:
\begin{equation}
\mathbf T(\tau) = \arg\max_{\{\rho_{ij}\}}\left(\sum_{i,j}\left(c_{ij}\log \rho_{ij}\right)\right)
\label{eq:7}
\end{equation}
subject to the row normalization constraint of the transition matrix $\sum_j \rho_{ij} = 1$ for all $i$. Since $\mathbf T(\tau)$ of the projected dynamics in the discretized state space satisfies detailed balance, it is common and reasonable to further impose the constraint:
\begin{equation}
\pi_i\rho_{ij} = \pi_j\rho_{ji}
\label{eq:8}
\end{equation}
for all $i$ and $j$ to the optimization problem in (\ref{eq:7}). Here, $\pi_i$,$\pi_j$ are elements of the stationary vector $\pi = [\pi_0,\pi_1,...,\pi_{L-1}]^\top$, which refers to the stationary probability distribution of finding the system in $S_i$ or $S_j$ at equilibrium, respectively. 

The optimization problem of (\ref{eq:7}) that satisfies both constraints can be solved by the following iterative algorithm\cite{bowman2009progress,trendelkamp2015estimation}:
\begin{equation}
\pi_i^{(k+1)} = \sum^n_{j=1} \frac{c_{ij}+c_{ji}}{\frac{c_i}{\pi_i^{(k)}}+\frac{c_j}{\pi_j^{(k)}}}
\label{eq:9}
\end{equation}
where one starts with an arbitrary non-zero guess of $\pi$ and computes each $\pi_i^{(k+1)}$ at the $(k+1)$ step from all entries of $\pi^{(k)}$ at the $k$th step of the iteration until convergence. Upon convergence, one can then compute all $\rho^*_{ij}$ by the following formula:
\begin{equation}
\rho^*_{ij} = \frac{(c_{ij}+c_{ji})\pi_j}{c_i\pi_j + c_j\pi_i}
\label{eq:10}
\end{equation}
where $c_i = \sum_j c_{ij}$. We refer to this optimization algorithm as the reversible estimator in this paper.

In contrast, if we know the exact value of $\pi$ of the underlying system prior to solving for $\mathbf T(\tau)$, we can improve the statistical efficiency and reduce the variance of our estimation of $\rho_{ij}$ by further fixing all $\pi_i$ and $\pi_j$ in (\ref{eq:8}) during optimization and thus limiting the parameter search space for $\rho^*_{ij}$. One can solve (\ref{eq:7}) with this additional $\pi$ constraint using the following iterative algorithm\cite{trendelkamp2015estimation}:
\begin{equation}
v_i^{(k+1)} = v_i^{(k)}\sum_j \frac{\pi_i(c_{ij}+c_{ji})}{v_j^{(k)}\pi_i  +  v_i^{(k)}\pi_j}
\label{eq:11}
\end{equation}
where $v = [v_1,v_2,...,v_L]^\top$ are the Lagrange multipliers. Similar to (9), one starts with an arbitrary guess of $v$ and iterates (11) until convergence. Upon convergence, one can compute all $\rho^*_{ij}$ by the following formula:
\begin{equation}
\rho^*_{ij} = \frac{(c_{ij}+c_{ji})\pi_j}{v_i\pi_j + v_j\pi_i}
\label{eq:12}
\end{equation}
We refer to this optimization algorithm as the $\pi$-constrained (reversible) estimator in this paper. In comparison to the reversible estimator, the $\pi$-constrained estimator is far less commonly used when estimating MSMs from unbiased simulations, mainly due to the reason that $\pi$ is often unknown a priori. Nevertheless, the $\pi$-constrained estimator is the core of our $\pi$-Girsanov method presented below, since $\pi$ can often be easily computed from well-converged enhanced sampling simulations.

Once we have obtained the final transition matrix $\mathbf T(\tau)$ using either (\ref{eq:10}) or (\ref{eq:12}), we can obtain the kinetic and thermodynamic properties of the system through the eigendecomposition of $\mathbf T(\tau)$:
\begin{equation}
\mathbf T(\tau)= \Psi \mathbf \Lambda \Psi^{-1}
\label{eq:13}
\end{equation}
where $\mathbf \Lambda = \text{diag}(\lambda_0,\lambda_1,...,\lambda_{L-1})$ is a diagonal matrix of eigenvalues $\lambda_i$ and $\Psi = [\psi_0,\psi_1,...,\psi_{L-1}]$ is a matrix that contains the right eigenvectors of $\mathbf T(\tau)$. The eigenvalues $\lambda_i$ are guaranteed to lie in the interval $[0,1]$ in descending order with the leading eigenvalue $\lambda_0$ always equal to 1. $\lambda_i$ can be used to compute the implied timescales (ITS) $t_i$ of a specific orthogonal dynamics mode of the system represented by its corresponding eigenvector $\psi_i$:
\begin{equation}
t_i = -\frac{\tau}{\log\lambda_i}
\label{eq:14}
\end{equation}
$t_i$ corresponds to the physical timescales of the system's dynamics modes and can be directly measured by spectroscopy experiment. Thus, it is of great significance that the MSM constructed can predict $t_i$ accurately. Notice for $\lambda_0 \equiv 1$, we have $t_0 \to \infty$ from (\ref{eq:14}) and the corresponding left eigenvector $\phi_0 \equiv \pi = [\pi_0,\pi_1,...,\pi_{L-1}]^\top$ represents the stationary distribution the Markov chain will converge into at an infinite timescale. Therefore, $\phi_0$ encodes the thermodynamics of the system.

\subsection{Girsanov Reweighting}
Now instead of running the MD simulations of the system at the original potential $U(q)$, we consider the case where we run the simulation at a biased potential $U^b(q) = U(q) +b (q)$ with a time-independent bias potential $b(q)$. One can employ Girsanov's theorem\cite{girsanov1960transforming} to obtain the unbiased estimate\cite{donati2017girsanov} of $\mathbf C(\tau)$ using a weighted version of (\ref{eq:5}) from paths sampled at the biased potential $U^b(x)$:
\begin{equation}
\begin{aligned}
c_{ij}
&= \int_\Gamma\frac{\pi(x_0)}{\pi^b(x_0)}\mathds{1}_{S_i}(x_{0})\pi^b(x_0)\int_\Omega \frac{\mu(\omega)}{\mu^b(\omega)}\mathds{1}_{S_j}(x_{\tau})\mu^b(\omega)\mathrm d\omega\mathrm dx_0\\
&= \int_\Gamma g(x_0)\mathds{1}_{S_i}(x_{0})\pi^b(x_0)\int_\Omega M(\omega)\mathds{1}_{S_j}(x_{\tau})\mu^b(\omega)\mathrm d\omega\mathrm dx_0\\
\end{aligned}
\label{eq:15}
\end{equation}
where $\pi^b(x_0)$ is the biased Boltzmann distribution of $x_0$ at $U^b(q)$ and $\mu^b(\omega)$ is the biased conditional probability density of observing a path $\omega$ given its starting point at $x_0$. We also define two reweighting factors in (\ref{eq:15}): $g(x_0):= \frac{\pi(x_0)}{\pi^b(x_0)}$ and $M(\omega) := \frac{\mu(\omega)}{\mu^b(\omega)}$. Throughout this paper, we also require that the absolute continuity condition is satisfied. That is, there is sufficient overlap between the unbiased and the biased ensemble, so that $g(x_0)$ and $M(\omega)$ are well defined for any $x_0$ and $\omega$ we have sampled from the biased ensemble. One can interpret $g(x_0)$ as the ratio between the probabilities of paths starting from these starting points at $U(q)$ and $U^b(q)$ whereas $M(\omega)$ can be interpreted as the ratio between the conditional probabilities of a specific path $\omega$ given a starting point $x_0$ at $U(q)$ and $U^b(q)$. The unbiased $\mathbf C(\tau)$ with paths $\omega$ from biased simulations at $U^b(q)$ can then be computed as\cite{donati2017girsanov}:
\begin{equation}
c_{ij} = \lim_{m\to\infty}\left[\frac{1}{m}\sum_{k=1}^{m}\left(g(x_{0}^{(k)})\mathds{1}_{S_i}(x_{0}^{(k)})M(\omega^{(k)})\mathds{1}_{S_j}(x_{\tau}^{(k)})\right)\right]
\label{eq:16}
\end{equation}
For $g(x_0)$, from its definition, one can see
\begin{equation}
g(x_0) = \frac{\pi(x_0)}{\pi^b(x_0)} = \frac{Z^b}{Z}\cdot e^{\frac{b(q_{0})}{k_BT}}
\label{eq:17}
\end{equation}
where $q_{0}$ is the Cartesian position coordinate of the path starting point and $Z^b$, $Z$ are the partition functions of the biased and unbiased ensemble, respectively. In the maximum likelihood estimation of $\mathbf T(\tau)$ using (\ref{eq:10}), the constant $Z^b/Z$ cancels\cite{donati2017girsanov}, and one can therefore simply let $g(x_0) = e^{{b(q_{0})}/({k_BT})}$. Nevertheless, it should be noticed that the formulation in (\ref{eq:15})-(\ref{eq:17}) inherently assumes the biased simulation has reached global equilibrium so that the initial points $x_0$ are distributed according to $\pi^b(x_0)$\cite{donati2017girsanov} (also see Appendix for details), which may be hard to achieve in certain scenarios. Finally, using the relationship in (\ref{eq:4}) (see Appendix for details), it was suggested that $M(\omega)$ can be computed as\cite{donati2017girsanov}
\begin{equation}
\begin{aligned}
M(\omega)  = \frac{\mu(\omega)}{\mu^b(\omega)}
&= \frac{\prod^{n_\tau-1}_{t=0}\rho(x_{t+1}|x_t)}{\prod^{n_\tau-1}_{t=0}\rho^b(x_{t+1}|x_t)}\\
&= \frac{\prod^{n_\tau-1}_{t=0}\prod^{3N}_{i=1}\exp\left(-\frac{{(\eta_{t,i})}^2}{2}\right)}{\prod^{n_\tau-1}_{t=0}\prod^{3N}_{i=1}\exp\left(-\frac{{(\eta_{t,i}^b)}^2}{2}\right)}\\
&= \prod^{n_\tau-1}_{t=0}\prod^{3N}_{i=1}\exp\left(-\eta_{t,i} \cdot\Delta\eta_{t,i} -\frac{1}{2}(\Delta\eta_{t,i})^2\right)
\end{aligned}
\label{eq:18}
\end{equation}
where $\Delta \eta_{t,i} = \eta_{t,i} - \eta^b_{t,i}$ is the $i$th random number difference at time $t$ and $n_\tau$ is the number of integration steps of $\omega$. For simulations numerically integrated by (\ref{eq:3}), $\Delta \eta_{t,i}$ can be computed as\cite{kieninger2023girsanov}
\begin{equation}
\Delta\eta_{t,i} = -\frac{(e^{-\xi\Delta t}+1)}{\sqrt {k_BT M_{ii}(1-e^{-2\xi\Delta t})}}\cdot\frac{\Delta t}{2}\cdot \frac{\partial b(q_{t+\frac{1}{2}\Delta t})}{\partial q_{t+\frac{1}{2}\Delta t,i}}
\label{eq:19}
\end{equation}
where $M_{ii}$ is the $i$th diagonal entry of the mass matrix $M$ and $\partial b(q_{t+\frac{1}{2}\Delta t})/\partial q_{t+\frac{1}{2}\Delta t,i}$ is the partial derivative of the bias potential evaluated on the system phase space coordinate at the intermediate integration step $q_{t+\frac{1}{2}\Delta t}$ with respect to the $i$th system coordinate $q_i$. Inserting (\ref{eq:17}), (\ref{eq:18}) and (\ref{eq:19}) into (\ref{eq:16}) one can then reweight the effect of $b(q)$ and obtain an unbiased estimate of $\mathbf C(\tau)$ for the maximum likelihood estimation of $\mathbf T(\tau)$, using either (\ref{eq:10}) or (\ref{eq:12}) similar to building MSMs from unbiased trajectories simulated at $U(q)$. Notice ($\ref{eq:18}$) states that $\eta_t$ and $\Delta \eta_t$ at every integration step for a path sample is required to compute $M(\omega)$ and clearly writing this high-dimensional vector at every integration step in practice is unrealistic. To tackle this, we applied an on-the-fly computation strategy for $M(\omega)$ from Ref.\citenum{schafer2024implementation}.

\subsection{The $\pi$-Girsanov Method}
Our goal is to extend the original Girsanov reweighting method above to the estimation of $\mathbf T(\tau)$ from biased trajectories generated from $J$ simulations with different time dependent or independent biased potentials $U^b_j(q,t) = U(q)+b_j(q,t)$ in the $j$th simulation. Such generalization allows us to construct MSMs from either multiensemble, non-equilibrium biased simulations, or even a mixture of them. 

From (\ref{eq:4}), (\ref{eq:18}) and (\ref{eq:19}) it is not hard to see that the computation of $M(\omega)$ can be easily generalized to such scenario since $\Delta \eta_{t,i}$ at each integration step only depends on the instantaneous value of $\partial b(q_{t+\frac{1}{2}\Delta t})/\partial q_{t+\frac{1}{2}\Delta t,i}$ at $t+\frac{1}{2}\Delta t$. 

The real difficulty in reweighting time-dependent or multiensemble biases lies in the computation of $g(x_0)$. For multiensemble simulations with time-independent biases, if we assume that each simulation has reached its global equilibrium in its own ensemble, then clearly, the $x_0$ of paths collected from all simulations do not follow any Boltzmann-type distributions but instead a mixture of these Boltzmann distributions from these different ensembles. In principle, one can compute the ratio between the unbiased Boltzmann distribution and this mixture distribution using multiensemble reweighting methods such as MBAR\cite{shirts2008statistically}. Yet, we will see in section \ref{sec:results} that this method is still practically sub-optimal. For simulations with time-dependent biases, the computation of $g(x_0)$ may, in theory, be computed from e.g. the Hummer-Szabo estimator\cite{hummer2001free,hummer2010free} based on Jarzynski's equality\cite{jarzynski1997nonequilibrium}. Yet, such estimator suffers from a high variance and may significantly reduce our statistical efficiency. It has also been shown in Ref.\citenum{donati2018girsanov} that $g(x_0)$ may be estimated directly using ($\ref{eq:17}$) from metadynamics simulations with a slow deposition rates such that the system was never far from equilibrium. Nevertheless, one cannot generalize such method to an arbitrary time-dependent $b_j(q,t)$. The complexity further increases if we consider combining multiple simulations with different $b_j(q,t)$.

Therefore, in our $\pi$-Girsanov method, we propose to let $g(x_0)=1$ and drop this term from the computation of $\mathbf C(\tau)$ in (\ref{eq:16}) completely. The resulting cross-correlation functions, $\hat c_{ij}$ of $\hat {\mathbf{C}}(\tau)$, computed as
\begin{equation}
\hat{c}_{ij}(\tau) = \lim_{m\to\infty}\left[\frac{1}{m}\sum_{k=1}^{m}\left(\mathds{1}_{S_i}(x_{0}^{(k)})M(\omega^{(k)})\mathds{1}_{S_j}(x_{\tau}^{(k)})\right)\right]
\label{eq:20}
\end{equation}
is apparently biased unless the distribution of $x_0$ from our biased simulations coincides with $\pi(x)$. Nevertheless, as long as we collected enough path samples, i.e. $m \to \infty$ and reweight with $M(\omega)$ correctly, the bias in $\hat c_{ij}(\tau)$ will only arise from the biased distribution of $x_0$. 

Interestingly, such scenarios have been seen in building MSMs with adaptive sampling\cite{bowman2010enhanced,kleiman2023adaptive} trajectories, where short unbiased trajectories are repetitively restarted from the least sampled region of the phase space to enhance phase space exploration. Both theoretical\cite{nuske2017markov,wu2017variational} and empirical\cite{wan2020adaptive} effort has been made on how one can reweight this bias effect. Specifically, Voelz et al.\cite{wan2020adaptive} empirically demonstrated that if the stationary distribution $\pi$ of $\mathbf T(\tau)$ is already known, one can use the $\pi$-constrained estimator in (\ref{eq:11}) and (\ref{eq:12}) to obtain an unbiased $\mathbf T(\tau)$ from the biased cross-correlation matrix $\hat {\mathbf C}(\tau)$ estimated from adaptive sampling trajectories with a biased distribution of $x_0$. 

On the other hand, a key difference between using biased trajectories with Girsanov reweighting and adaptive sampling trajectories is that, for adaptive sampling trajectories, it is, in general, very challenging to obtain the unbiased equilibrium distribution $\pi(x)$ or the stationary vector $\pi$ before the computation of $\mathbf T(\tau)$, whereas one can easily obtain $\pi(x)$ from most popular enhanced sampling algorithms (see below) with high statistical confidence using existing thermodynamic reweighting methods. The strategy of using biased simulation trajectories is therefore much more practically applicable. Moreover, since $\hat c_{ij}$ has already been estimated with a biased distribution of $x_0$, we can further drop the assumption that the simulation has reached global equilibration in the biased ensemble required by the original Girsanov reweighting formulation mentioned above. We postulate that the bias introduced from either non-equilibrium sampling or setting $g(x_0)= 1$ can be simultaneously removed with the $\pi$-constrained estimator of $\mathbf T(\tau)$. To further justify our method, in Appendix, we show that setting $g(x_0)=1$ can also be interpreted as merely assigning a different weight to each path sample, where the optimization goal in (\ref{eq:7}) remains meaningful even if we substitute $\hat{c}_{ij}$ into (\ref{eq:7}). It follows that the introduced $\pi$ constraint can be interpreted as a physical regularization that improves the numerical optimization, so that every estimated $\rho_{ij}$ in $\mathbf T(\tau)$ from data is constrained to be closer to its true unbiased value. 

Following the reasoning above, we propose the following protocol for the estimation of $\mathbf T(\tau)$ from biased trajectories generated at multiple different $U_j^b(q,t)$, which we refer to as the $\pi$-Girsanov method:
\begin{enumerate}
  \item Collect path samples $\omega$ from simulations at an arbitrary number of different $U^b_j(q,t)$ with an underdamped Langevin dynamics integrator e.g. ABOBA with known formulations of $\Delta \eta_{t,i}$ and $M(\omega)$.
  \item Compute the reweighted but biased cross-correlation matrix $\hat{\mathbf C}(\tau)$ with (\ref{eq:20}).
  \item Compute the unbiased Boltzmann distribution $\pi(x)$ with thermodynamic reweighting and then the stationary vector $\pi$ (see below).
  \item Estimate the unbiased $\mathbf T(\tau)$ from $\hat{\mathbf C}(\tau)$ using the $\pi$-constrained estimator from (\ref{eq:11}) and (\ref{eq:12}).
\end{enumerate}
Notice this method also generalizes to scenario where the simulations are performed at a single biased potential $U^b(q)$ with a time-independent bias $b(q)$ which the original Girsanov reweighting method applies. We will see later in section {\ref{sec:results}} that the $\pi$-Girsanov method not only generalizes but also performs better in such scenario. 

\section{Methods}
\subsection{Enhanced Sampling Simulations}
We first tested our $\pi$-Girsanov method on enhanced sampling trajectories of three different analytical potentials: the 1D double well potential, the 1D Prinz potential\cite{prinz2011markov} and the 2D M\"{u}ller-Brown potential \cite{muller1979location}. For each analytical potential, biased trajectories from build-up and rerun (well-tempered) metadynamics, umbrella sampling and steered MD simulations were harvested and used to build MSMs to test the $\pi$-Girsanov method and compare with the previous Girsanov reweighting method. Here, the build-up metadynamics refer to biased simulations running with a single time-dependent bias potential $b^{\text{MetaD}}(q,t)$ defined as\cite{bussi2020using}:
\begin{equation}
b^{\text{MetaD}}(q,t) = \sum^{t'<t}_{t'=\tau_G,2\tau_G,...} \left[h_{t'}\exp\left(-\frac{(r(q)-r_{t'})^2}{2\sigma^2}\right)\right]
\label{eq:21}
\end{equation}
where $r(q) \in \mathbb R$ is a one-dimensional function of $q$, which is often referred to as the collective variable (CV), or reaction coordinate (RC), $\sigma$ is the bandwidth of Gaussian kernel functions defined by the user, $r_{t'}$ is the CV value at time $t'$, $\tau_G$ is the time interval between Gaussian kernel depositions and $h_{t'}$ is the height of the Gaussian kernel at time $t'$ which decays over time:
\begin{equation}
h_{t'} = h_0\exp\left(-\frac{b^{\text{MetaD}}(q,t')}{\Delta T}\right)
\label{eq:22}
\end{equation}
where $h_0$ is the user-defined initial Gaussian height and $\Delta T$ is related to the bias factor, $\gamma$, which is another user-defined parameter that controls the magnitude of this decay, via $\gamma = (T+\Delta T)/T$. At the long time limit $t\to\infty$, the bias potential in (\ref{eq:22}) will converge into\cite{dama2014well}
\begin{equation}
b^{\text{MetaD}}_{t\to\infty}(q) = \left(\frac{1}{\gamma}-1\right)k_BT\log \pi(r(q))
\label{eq:23}
\end{equation}
where $\pi(r(q))$ is the equilibrium distribution projected on $r(q)$ defined as $\pi(r(q))\propto \int \delta(r-r(q))\pi(x)\mathrm dx$ with $\delta(r-r(q))$ being the Dirac delta function centered at $r(q)$. We refer to biased simulations running with the bias potential in (23) as the rerun metadynamics simulation, as one may first build this bias potential from build-up simulations with (21) and then perform a rerun with the converged potential in (23) fixed. Notice this is the original strategy proposed to combine the original Girsanov reweighting method with metadynamics in Ref.\citenum{donati2018girsanov} and we are following their naming convention.

Umbrella sampling performs multiensemble biased simulations with time-independent harmonic bias potentials. The bias potential $b_j^{\text{US}} (q)$ of the $j$th ensemble of umbrella sampling is defined as:
\begin{equation}
b_j^{\text{US}} (q)= \frac{1}{2}\kappa(r(q)-\bar r_{j})^2
\label{eq:24}
\end{equation}
where $\kappa$ is the spring constant that controls the stiffness of the harmonic potential and $\bar r_j$ is a constant that controls the minimum of $b_j^{\text{US}} (q)$ for the $j$th simulation. Since each of these simulations will constrain the trajectory locally around $\bar r_j$, simulations at $b_j^{\text{US}} (q)$ with multiple equally spaced $\bar r_j$ are required to cover the CV space of interest.

Steered MD performs non-equilibrium simulations with the bias potential $b^{\text{SMD}}(q,t)$ defined as a harmonic potential with a moving center $\bar r(t)$ at a constant velocity $\dot r = \mathrm dr/\mathrm dt$:
\begin{equation}
b^{\text{SMD}}(q,t) = \frac{1}{2}\kappa(r(q)-\bar r(t))^2
\label{eq:25}
\end{equation}
where $\bar r(t) = \bar r(0) - \dot rt$ is the location of the minimum of the $b^{\text{SMD}}(q,t)$ at time $t$. Physically, this means gradually pulling the system in the CV space from an initial CV coordinate $\bar r(0)$ to another CV coordinate end point at the end of the simulation. 

For each analytical potential, we also perform long unbiased simulations to build MSMs as our reference. For 1D systems, $r(q)$ is simply defined as the only coordinate. For the 2D M\"{u}ller-Brown potential, we use one of the unbiased simulation trajectories as training data to obtain an optimal curvilinear 1D RC as $r(q)$ with our previously developed algorithm flow matching for reaction coordinates (FMRC) \cite{zhang2024flow}. 

We later applied the $\pi$-Girsanov method to build MSMs from enhanced sampling simulations of a model biomolecular systems: alanine dipeptide in implicit solvent. We performed two parallel build-up metadynamics simulations, each simulation with one of the dihedral angles ($\phi$ and $\psi$) as a 1D CV. We also performed long unbiased simulations of alanine dipeptide to build MSM as a reference.

\subsection{Computation of the Stationary Distribution from Enhanced Sampling Simulations}
For rerun metadynamics, we computed the unbiased Boltzmann distribution $\pi(x)$ with sample weight $w(x) = \exp(b^{\text{MetaD}}_{t\to\infty}(q)/(k_BT))$:
\begin{equation}
\pi(x) = \frac{\int w(x)\delta (x-x')\mathrm dx}{\int w(x)\mathrm dx}
\label{eq:26}
\end{equation}
where $\delta(x-x')$ is a Dirac delta function centered at $x'$. 

For build-up metadynamics, we used the reweighting method from \cite{branduardi2012metadynamics} with the assumption that the $b^{\text{MetaD}}(q,t)$ is pseudostatic. This method computes sample weights as $w(x)=\exp(b^{\text{MetaD}}(q,t_{\text{end}})/(k_BT))$ where $b^{\text{MetaD}}(q,t_{\text{end}})$ is the bias potential at the end of the simulation and computes $\pi(x)$ using the same formula in (\ref{eq:26}) with all samples. 

For umbrella sampling, we used MBAR\cite{shirts2008statistically} to compute the sample weights $w(x)$ and apply (\ref{eq:26}) to compute $\pi(x)$.

For steered MD simulations, due to the large variance of the Jarzynski's equality based estimators, we did not obtain $\pi(x)$ from our steered MD trajectories. Instead, for 1D potentials, we used the analytical $\pi(x)$ for testing our $\pi$-Girsanov method. For M\"{u}ller-Brown potential, we used the empirical distribution from the reference unbiased simulations for $\pi(r(q))$.

For the parallel build-up metadynamics simulations of alanine dipeptide, we first follow the assumption above that $b^{\text{MetaD}}(q,t)$ for each of the two simulations is pseudostatic. We then regard each simulation was carried out at the potential $b^{\text{MetaD}}(q,t_{\text{end}})$ and compute the bias potential values at each ensemble for all samples collected. We then substitute these values into the MBAR equations to compute $w(x)$ for each sample and use (\ref{eq:26}) to estimate $\pi(x)$.

After obtaining $\pi(x)$, we use the following formula to compute the stationary distribution $\pi$ for the estimation of $\mathbf T(\tau)$ with the $\pi$-constrained estimator:
\begin{equation}
\pi_i = \int_{x\in S_i} \pi(x)\mathrm dx
\label{eq:27}
\end{equation}

Notice here in (\ref{eq:26}) and (\ref{eq:27}), we have defined the stationary distribution in terms of the phase space coordinate $x$ directly for notation simplicity. In all of our numerical experiments, our discretization of the phase space is performed directly in the Cartesian coordinate space (1D system) or a reduced collective variable space (higher-dimensional system) only. As a result, we can obtain $\pi_i$ for all $S_i$ by only evaluating and integrating the marginal stationary distribution over the reduced space.

\subsection{MSM Construction and Evaluation}
To validate our $\pi$-Girsanov method and compare it with existing methods, for each analytical system and for each type of enhanced sampling trajectories generated from each of the system, we built MSMs with four different methods:
\begin{enumerate}
    \item Build a reference MSM with long unbiased trajectories using the reversible estimator in (\ref{eq:9}) and (\ref{eq:10}) with standard procedures.
    \item Let $g(x_0) = w(x_0)$ from (\ref{eq:26}), compute $\mathbf C(\tau)$ using (\ref{eq:16}) with $M(\omega)$ from (\ref{eq:18}) and build the MSM with the reversible estimator. This is the previous Girsanov reweighting method with the most straightforward extension to build-up metadynamics and umbrella sampling simulations. For steered MD simulations, we did not include this method for comparison.
    \item Compute $\hat{\mathbf C}(\tau)$ with (\ref{eq:20}) (equivalent to using (\ref{eq:16}) with $g(x_0)=1$) and build the MSM with the reversible estimator. This is a reference to evaluate the effect of ignoring $g(x_0)$ if we do not use a $\pi$-constraint for $\mathbf T(\tau)$ estimation.
    \item Compute $\hat{\mathbf C}(\tau)$ with (\ref{eq:20}), estimate $\pi$ with (\ref{eq:26}) and (\ref{eq:27}) (or use analytical $\pi$ for steered MD) and build the MSM with the $\pi$-constrained estimator in (\ref{eq:11}) and (\ref{eq:12}). This is our $\pi$-Girsanov method. 
\end{enumerate}

Furthermore, for each set of enhanced-sampling trajectories from a given system, we first performed an implied timescale (ITS) analysis. Ideally, this requires that the ITS \(t_i\) associated with the dominant slow processes form a plateau as a function of the lag time $\tau$ and remain approximately constant over the interval $[\tau_{\mathrm{Markov}}, t_i]$, where $\tau_{\mathrm{Markov}}$ marks the onset of the plateau. In principle, any $\tau$ within this interval is a valid choice for constructing the final MSM and recovering accurate thermodynamic and kinetic properties. However, in path reweighting, the variance of $M(\omega)$ increases rapidly with $\tau$ \cite{donati2017girsanov,schafer2024implementation}, which generally precludes the use of $\tau$ close to $t_i$ values for MSM construction. Therefore, in this work, we instead require that \(t_i\) remain approximately constant over a shorter interval beginning at $\tau_{\mathrm{Markov}}$, over which the variance remains well controlled. A value of $\tau$ selected from this interval was then used to construct both the final path-reweighted MSMs and the corresponding reference MSM for that set of trajectories.

We then eigendecomposed all $\mathbf T(\tau)$ we obtained and compared their eigenvectors. For each MSM from biased simulations of the analytical systems, we repeated the corresponding simulations for 10 times to obtain an evaluation of the variance (through standard deviations of these repeats) for our estimation of the ITS and eigenvectors as error bars. 

For the application in alanine dipeptide, we only built a reference MSM with long unbiased trajectories using the reversible estimator and an MSM from a single set of parallel build-up metadynamics trajectories with the $\pi$-Girsanov method without evaluating the variance of our estimation.

\subsection{Details in Simulation, Modeling and Software}

All simulations in this study were performed with the software package OpenMM 8.2.0\cite{eastman2023openmm} along with a custom version of openmmtools available at: https://github.com/anyschaefer/openmmtools. Specifically, for all simulations we propagated the underdamped Langevin dynamics defined in (\ref{eq:1}) numerically with the ABOBA integrator implementation under the class \texttt{LangvevinSplittingGirsanov} in this custom openmmtools plugin\cite{schafer2024implementation}, which allows on-the-fly computation of $M(\omega)$.

The unbiased potential energy function $U^{\text{DW}}(q_1)$ of the 1D double well potential along the system's only coordinate $q_1$ is defined as:
\begin{equation}
U^{\text{DW}}(q_1) = -50e^{-\frac{(q_1+0.5)^2}{0.25}}-50e^{-\frac{(q_1-0.5)^2}{0.25}}
\label{eq:28}
\end{equation}

The unbiased potential energy function $U^{\text{Prinz}}(q_1)$ of the 1D Prinz potential along the system's only coordinate $q_1$ is defined as:
\begin{equation}
U(q_1) = 20q_1^8+ 16e^{-80q_1^2} + 4e^{-80(q_1-0.5)^2} + 10e^{-40(q_1+0.5)^2}
\label{eq:29}
\end{equation}

The unbiased potential energy function $U^{\text{MB}}(q_1,q_2)$ of the 2D M\"{u}ller-Brown potential along the two coordinates $q_1$ and $q_2$ is defined as:
\begin{equation}
U^{\text{MB}}(q_1,q_2) = \sum_{i=1,2,3,4} A_i \exp(a_i(q_1-\bar q_{1,i})^2-b_i(q_1-\bar q_{1,i})(q_2-\bar q_{2,i})-c_i(q_2-\bar q_{2,i})^2)
\label{eq:30}
\end{equation}
where we have $A=[-40,-20,-34,3]$, $a=[-1,-1,-6.5,0,7]$, $b=[0,0,11,0.6]$, $c=[-10,-10,-6.5,0.7]$, $\bar q_1 = [1,0,-0.5,-1]$ and $\bar q_2 = [0,0.5,1.5,1]$.

All analytical potential simulations were performed on a single particle with a mass equal to 1 Dalton. The simulation parameters for all analytical potentials are as follows: $T = 298.15\text{K}$, $\xi = 10 \text{ps}^{-1}$, $\Delta t = 5\text{fs}$.

For each analytical system, 10 independent unbiased simulations were performed for 100 ns each as reference. 10 independent rerun metadynamics simulations were performed for 10 ns each with $\gamma = 2$. 10 independent build-up simulations were performed for 10 ns each with $\gamma=2$, $\sigma=0.1$, $h_0=1.2 \text{kJ/mol}$ and $\tau_G=0.1\text{ps}$. For umbrella sampling, we use $\kappa = 100$ and 50 umbrella windows evenly distributed across $r(q)$ and simulated each window for $0.2\text{ns}$. For steered MD simulations, the system was slowly pulled from one end of $r(q)$ to the other end at $\dot r = 1.6\times10^{-3}$ (double well and Prinz) or $\dot r = 1.7\times10^{-3}$ (M\"{u}ller-Brown) and is pulled in reverse direction whenever it hits one of the endpoint. Such steered MD simulation was performed for a total of 10ns for each system.

The alanine dipeptide system is modeled with the AMBERSB99-ILDN force field\cite{lindorff2010improved} for the peptide and the OBC model \cite{onufriev2004exploring} for implicit solvent. The simulation parameters for alanine dipeptide are as follows: $T = 310\text{K}$, $\xi = 100 \text{ps}^{-1}$, $\Delta t = 2\text{fs}$. For reference simulations, 5 independent unbiased simulations were performed for 1 $\mu$s. The build-up metadynamics simulations were performed for 50 ns each with $\gamma=2$, $\sigma=0.1$, $h_0=1.2 \text{kJ/mol}$ and $\tau_G=1\text{ps}$.

All analytical potential and alanine dipeptide trajectories above are recorded at an interval $\Delta t^{\text{int}}=0.1\text{ps}$  except the reference simulation of alanine dipeptide, which is recorded at an interval $\Delta t^{\text{int}}=1\text{ps}$. We summarized the MD simulations performed in this study in Table~\ref{tab:sim_setup}.

\begin{table}[t]
\centering
\resizebox{0.9\textwidth}{!}{%
\begin{tabular}{|l|ccc|c|}
\hline
System                  & \multicolumn{1}{c|}{1D Double Well Potential} & \multicolumn{1}{c|}{1D Prinz Potential} & 2D M\"uller-Brown Potential                       & Alanine Dipeptide                                                                                           \\ \hline
Unbiased MD (Reference) & \multicolumn{3}{c|}{$10$ parallel simulations $\times$ $10$ ns}                                                                              & $1 \times 1$ $\mu$s                                                                                         \\ \hline
Rerun Metadynamics      & \multicolumn{2}{c|}{$10$ repeats $\times $ $1$ ns}                                         & $10$ repeats $\times$ $2.5$ ns                    & -                                                                                                           \\ \hline
Build-up Metadynamics   & \multicolumn{2}{c|}{$10$ repeats $\times $ $1$ ns}                                         & $10$ repeats $\times$ $2.5$ ns                    & \begin{tabular}[c]{@{}c@{}}$1 \times 50$ ns ($\phi$ as CV)\\ $1 \times 50$ ns ($\psi$ as CV)\end{tabular}                                                                                                           \\ \hline
Umbrella Sampling       & \multicolumn{2}{c|}{$10$ repeats $\times $ $50 \times 20$ ps}                              & $10$ repeats $\times$ $50$ windows $\times$ $50$ ps & -                                                                                                           \\ \hline
Steered MD              & \multicolumn{2}{c|}{$10$ repeats $\times $ $1$ ns}                                         & $10$ repeats $\times$ $2.5$ ns                    & -                                                                                                           \\ \hline
\end{tabular}%
}
\caption{A summary of simulations for the different systems performed in this study. A dash (-) indicates that no simulation was performed for this particular simulation method.}
\label{tab:sim_setup}
\end{table}

The FMRC model for the optimal RC of the 2D M\"{u}ller-Brown potential is trained on the first unbiased trajectory as follows: the system coordinates $q_1$ and $q_2$ were first pre-processed by a method known as the time-lagged independent component analysis (TICA) \cite{perez2013identification} with $\tau = \Delta t^{\text{int}}$. All TICA eigenvectors are then used as the input for the encoder of FMRC. The encoder and decoders are both fully connected neural networks composed of an input layer, an output layer and a hidden layer of 16 neurons with RELU\cite{agarap2018deep} as activation functions. Other FMRC parameters are the same as our previous work in Ref. \citenum{zhang2024flow}.

All MSMs were built and analyzed using the software package Deeptime 0.4.5 \cite{hoffmann2021deeptime}. For MSM construction from the trajectories of all analytical potentials, we discretized the CV space into 25 grid cells equally spaced along $r(q)$ as the microstates of our systems. One appropriate $\tau$ is chosen independently for each $\mathbf T(\tau)$ from each set of enhanced sampling trajectories based on the ITS plot as mentioned above. For 1D double well potential, $\tau = 10\Delta t^{\text{int}}$ are chosen for rerun and build-up metadynamics, $\tau = 4\Delta t^{\text{int}}$ are chosen for umbrella sampling and $\tau = 3\Delta t^{\text{int}}$ are chosen for steered MD. For 1D Prinz potential, $\tau = 8\Delta t^{\text{int}}$ are chosen for rerun and build-up metadynamics, $\tau = 1\Delta t^{\text{int}}$ are chosen for umbrella sampling and $\tau = 5\Delta t^{\text{int}}$ are chosen for steered MD. For 2D M\"{u}ller-Brown potential, $\tau = 5\Delta t^{\text{int}}$ are chosen for rerun and build-up metadynamics, $\tau = 1\Delta t^{\text{int}}$ are chosen for umbrella sampling and steered MD.

For alanine dipeptide, MSM was built as follows: we discretized the $\phi$-$\psi$ 2D space with $50\times50$ regular 2D grids. We then use $\tau = 10\Delta t^{\text{int}}$ for the estimation of $\mathbf T(\tau)$ from both unbiased and biased trajectories.

All the simulation and analysis scripts required to replicate the study are available at: https://github.com/Mingyuan00/Pi-Girsanov.

\section{Results}
\label{sec:results}

\begin{figure}[h]
    \centering
    \includegraphics[scale=0.05]{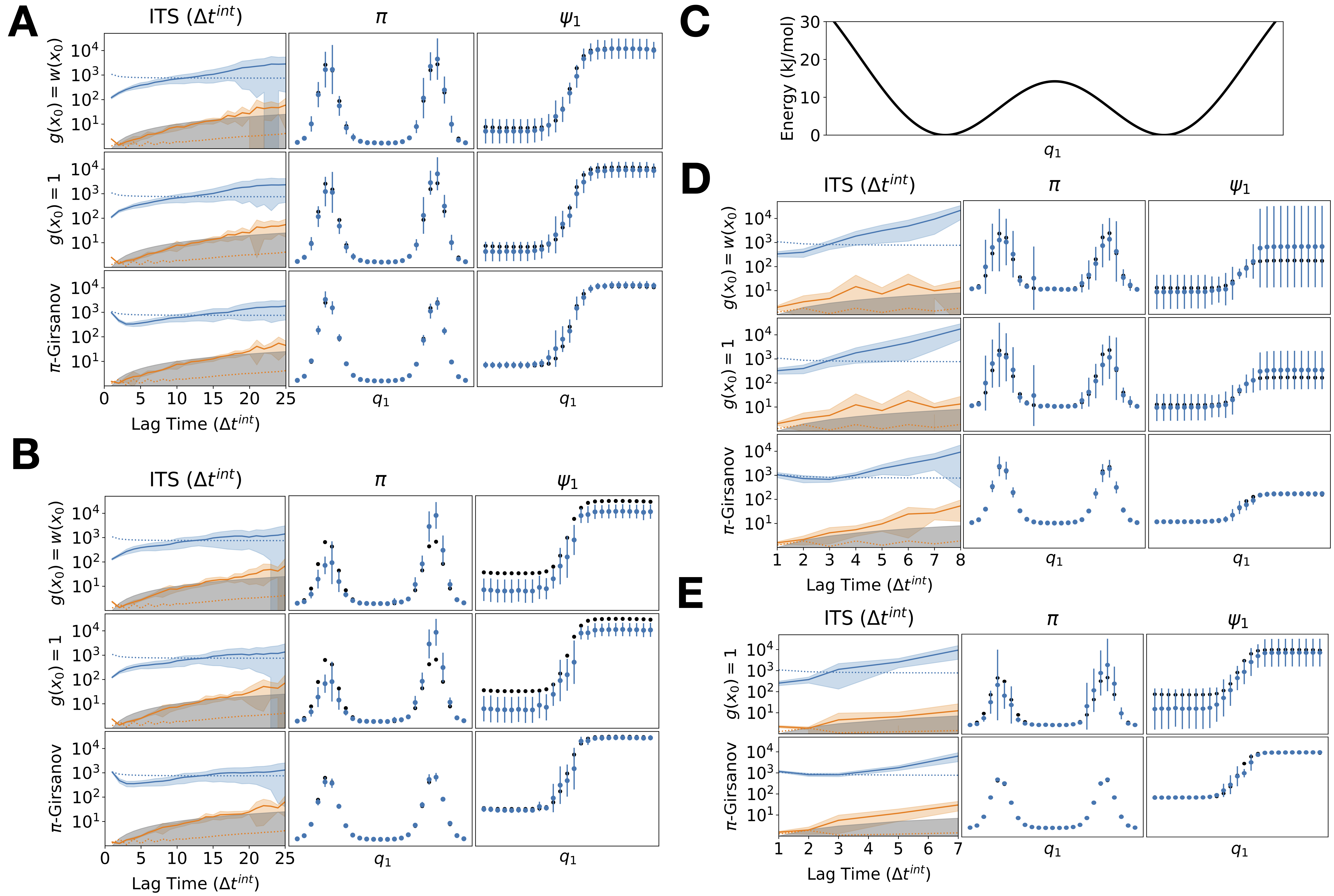}
    \caption{Summary of results for 1D double well potential simulations: (A) The ITS, the stationary distribution $\pi$ and the eigenvector $\psi_1$ of $\mathbf T(\tau)$ reweighted with different methods from the rerun metadynamics trajectories. (B) The ITS, the stationary distribution $\pi$ and the eigenvector $\psi_1$ of $\mathbf T(\tau)$ reweighted with different methods from the build-up metadynamics trajectories. (C) The energy landscape of the system. (D) The ITS, the stationary distribution $\pi$ and the eigenvector $\psi_1$ of $\mathbf T(\tau)$ reweighted with different methods from the umbrella sampling trajectories. (E) The ITS, the stationary distribution $\pi$ and the eigenvector $\psi_1$ of $\mathbf T(\tau)$ reweighted with different methods from the steered MD trajectories. For all ITS plots, we use dotted lines to represent the reference values obtained from the reference MSM, solid lines to represent the values from the reweighted MSM and shade to represent the variance. For all eigenvector plots, we use black dots to represent the reference values and blue dots along with error bars to represent the reweighted values and their variance. For each column of each subplots, the y-axis scale is set to uniform so comparison can be made.}
    \label{fig:1}
\end{figure}

\subsection{1D Double Well Potential}
We summarize the results of applying the $\pi$-Girsanov method and the other methods for comparison to the biased trajectories from different enhanced sampling simulations of the 1D double well potential in Figure 1. Since there are only two metastable states in the system (Figure 1C), we only compare the stationary vector $\pi$, the system's major slow mode $\psi_1$ and its associated time scale $t_1$ computed with each method.

We first performed the ITS analysis for MSMs built with different methods for each set of biased trajectories from different enhanced sampling simulations. For all MSMs built from different reweighting methods with different set of biased trajectories, we were able to recover an accurate estimation of $t_1$ around $10^3 \Delta t^{\text{int}}$ aligned with the reference $t_1$ at a certain $\tau$. Nevertheless, for rerun or build-up metadynamics simulations, we find that estimated $t_1$ generally converges to a plateau region, but the variance increases rapidly across this region. For trajectories from either umbrella sampling or steered MD, we cannot even converge the iterative algorithm for the MLE of $\mathbf T(\tau)$ with a $\tau$ larger than $10 \Delta t^{\text{int}}$. However, one can see in both scenarios, $t_1$ of the MSMs computed with the $\pi$-Girsanov method tend to have a smaller variance and fewer deviations from the reference value compared to other methods.

For rerun metadynamics, build-up metadynamics and umbrella sampling (Figure 1A, 1B and 1D), direct $\pi$ estimation from samples using (26) and (27) with $w(x)$ from thermodynamic reweighting methods is able to obtain $\pi$ with a minimal deviation and variance with the limited biased trajectories. Using this $\pi$ as a constraint for $\mathbf T(\tau)$ estimation, the $\pi$-Girsanov method is able to obtain consistent $\psi_1$ from repeated experiments that aligns well with the reference $\psi_1$ for all three enhanced sampling methods, with only some slight discrepancies and increment in variance at the transition region. In contrast, $\pi$ and $\psi_1$ estimated from the MSMs using previous Girsanov reweighting methods, either computing or ignoring $g(x_0)$, converge significantly slower with a higher variance. Even for the rerun metadynamics case where the method was originally developed to tackle and performs the best, there is a higher variance across repeated experiments in comparison to the $\pi$-Girsanov method. If we attempt to use $w(x_0)$ as our $g(x_0)$ or simply ignoring $g(x_0)$ for reweighting build-up metadynamics or umbrella sampling trajectories with the original method, the resulting $\pi$ and $\psi_1$ will suffer from a even larger variance (umbrella sampling), or bias (build-up metadynamics). 

For steered MD trajectories (Figure 1E) where $g(x_0)$ is hard to obtain, if an accurate $\pi$ (in this case, analytical) can be provided, the $\pi$-Girsanov method is also able to recover $\psi_1$ that is very close to reference. In contrast, if we simply ignore the $g(x_0)$ factor and estimate $\mathbf T(\tau)$ without the $\pi$ constraint, the resulting $\pi$ and $\psi_1$ will similarly possess a high variance and bias.

\subsection{1D Prinz Potential}
\begin{figure}[h]
    \centering
    \includegraphics[scale=0.05]{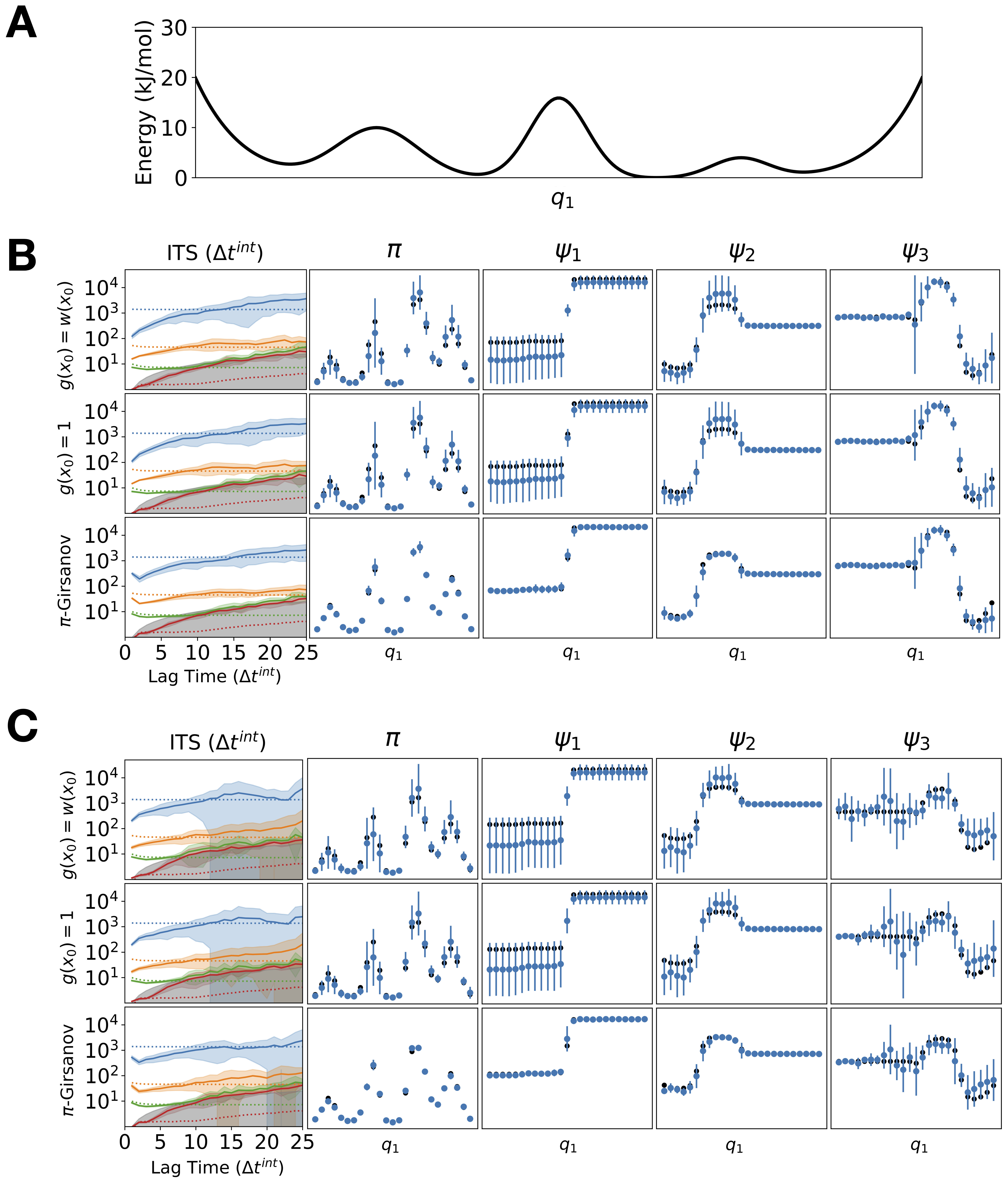}
    \caption{Summary of results for 1D Prinz potential rerun and build-up metadynamics simulations: (A) The energy landscape of the system. (B) The ITS, the stationary distribution $\pi$ and the leading eigenvectors $\psi_1$, $\psi_2$, $\psi_3$ of $\mathbf T(\tau)$ reweighted with different methods from the rerun metadynamics trajectories. (C) The ITS, the stationary distribution $\pi$ and the leading eigenvectors $\psi_1$, $\psi_2$, $\psi_3$ of $\mathbf T(\tau)$ reweighted with different methods from the build-up metadynamics trajectories. For all ITS plots, we use dotted lines to represent the reference values obtained from the reference MSM, solid lines to represent the values from the reweighted MSM and shade to represent the variance. For all eigenvector plots, we use black dots to represent the reference values and blue dots along with error bars to represent the reweighted values and their variance. For each column of each subplots, the y-axis scale is set to uniform so comparison can be made.}
    \label{fig:2}
\end{figure}

\begin{figure}[h]
    \centering
    \includegraphics[scale=0.05]{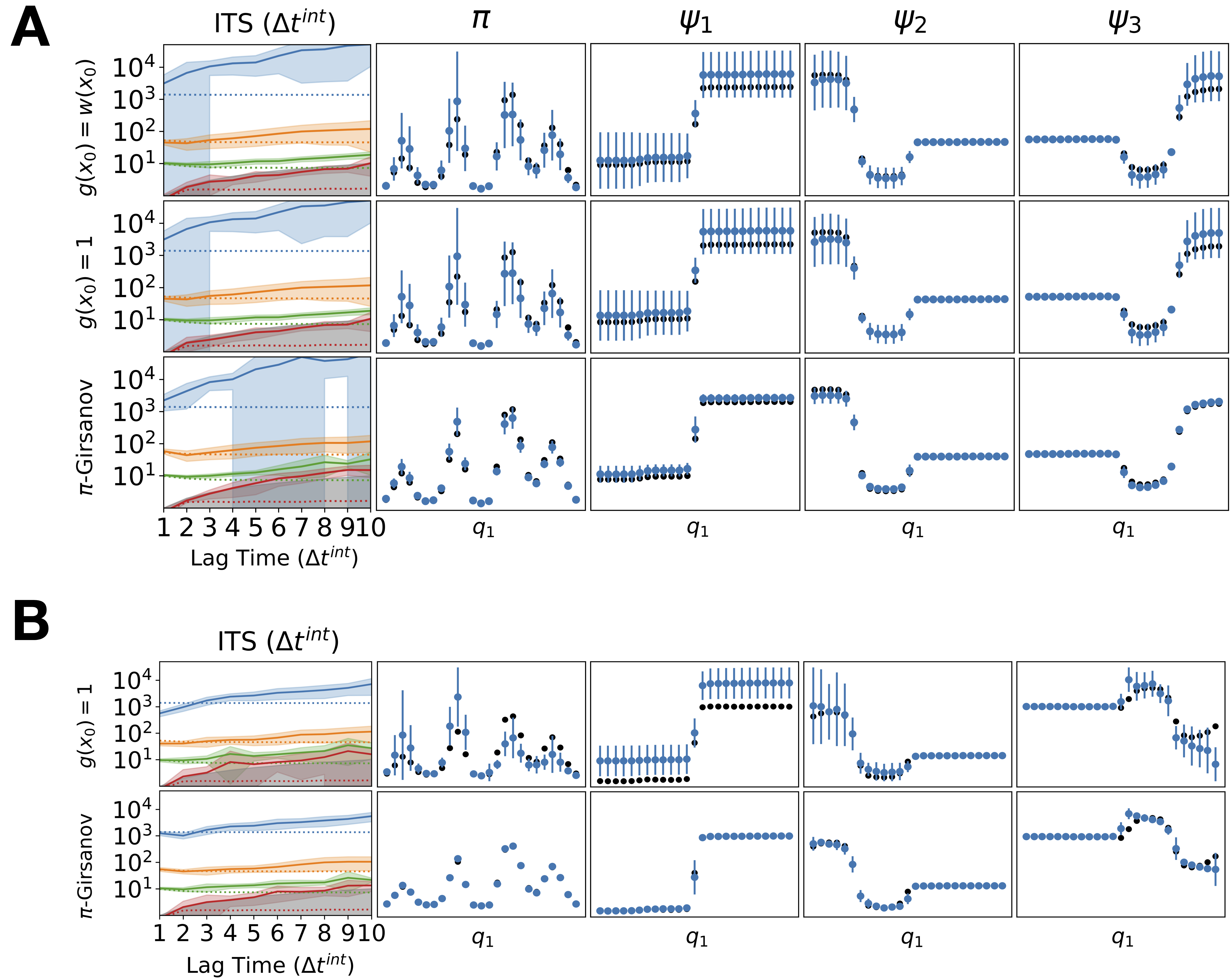}
    \caption{Summary of results for 1D Prinz potential umbrella sampling and steered MD simulations: (A) The ITS, the stationary distribution $\pi$ and the leading eigenvectors $\psi_1$, $\psi_2$, $\psi_3$ of $\mathbf T(\tau)$ reweighted with different methods from the umbrella sampling trajectories. (B) The ITS, the stationary distribution $\pi$ and the leading eigenvectors $\psi_1$, $\psi_2$, $\psi_3$ of $\mathbf T(\tau)$ reweighted with different methods from the steered MD trajectories. For all ITS plots, we use dotted lines to represent the reference values obtained from the reference MSM, solid lines to represent the values from the reweighted MSM and shade to represent the variance. For all eigenvector plots, we use black dots to represent the reference values and blue dots along with error bars to represent the reweighted values and their variance. For each column of each subplots, the y-axis scale is set to uniform so comparison can be made.}
    \label{fig:3}
\end{figure}

Motivated by the results above, we then proceeded to test our $\pi$-Girsanov method for the 1D Prinz potential from Ref.\citenum{prinz2011markov} and compare its performance with the other methods. Notice we have upscaled the energy barriers to a similar value with the 1D double well potential (Figure 2A) in comparison to Ref.\citenum{prinz2011markov} to ensure that this is a more challenging case. Similar to above, we also generated biased trajectories with four different enhanced sampling methods. Since there are four metastable states in the system, we focus on the estimation of $\pi$, $\psi_1$, $\psi_2$ and $\psi_3$ along with their ITS $t_1$, $t_2$ and $t_3$ by different methods. These results are summarized in Figure 2 and Figure 3. 

With the increased complexity in the system, the variance of the ITS estimation significantly increases across all reweighting methods and enhanced sampling methods. The estimation of the ITS has become particularly challenging for umbrella sampling simulations (Figure 3A), as none of the reweighting methods, including $\pi$-Girsanov, could obtain $t_1$ without a high variance, even with a small $\tau$. Similar behaviours were also observed for estimating $t_1$ and $t_2$ from build-up metadynamics simulations (Figure 2C) with the original Girsanov method at large $\tau$, which can be improved using the $\pi$-Girsanov method. Overall, the limitation of variance in $M(\omega)$ persists to be an issue for the ITS with a large $\tau$ and $\pi$-Girsanov methods still perform better in most cases in comparison to the original method.

Similar to above, we again see where applicable, thermodynamic reweighting methods are able to provide an estimation of $\pi$ with less variance than the original Girsanov reweighting using the same amount of trajectories for all enhanced sampling methods. Using this $\pi$ as constraint for $\mathbf T(\tau)$ estimation, the $\pi$-Girsanov method again outperforms the other methods in terms of the quality of the eigenvectors obtained. We also see the original Girsanov reweighting method starts to struggle in obtaining accurate $\psi_1$ with the rerun metadynamics trajectories as the system complexity increases.

Again, ideally if one is provided with an accurate estimate of $\pi$, biased trajectories generated by steered MD simulations can be used to estimate an accurate $\mathbf T(\tau)$ with the $\pi$-Girsanov method. In contrast, in this system if we ignore the starting point weights $g(x_0)$, we can see that a significantly biased $\pi$ estimate from the eigenvectors of the resulting $\mathbf T(\tau)$ will be obtained, likely due to the starting point distribution of all paths is significantly deviated from $\pi(x)$ due to the non-equilibrium pulling force $b^{\text{SMD}}(q,t)$.

\subsection{2D M\"{u}ller-Brown Potential}

\begin{figure}[h]
    \centering
    \includegraphics[scale=0.45]{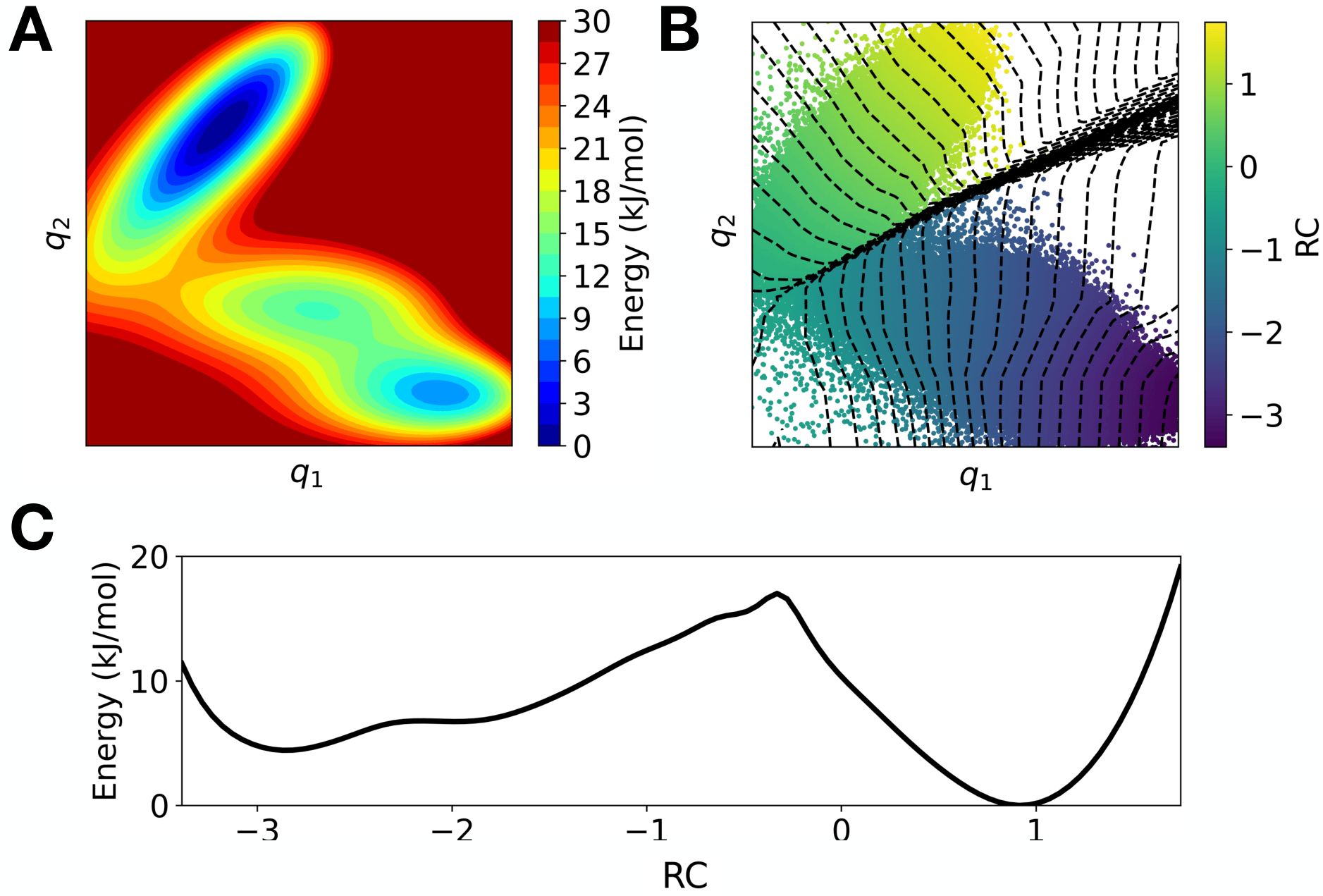}
    \caption{Dimensionality reduction of the 2D M\"{u}ller-Brown potential. (A) The energy landscape of the potential. (B) The scatterplot of the sample coordinates used in FMRC training. The resulting RC is shown as colorbar and its isolines are plotted as black dashed lines. (C) The projected free energy surface computed from the training data along the RC.}
    \label{fig:4}
\end{figure}

\begin{figure}[h]
    \centering
    \includegraphics[scale=0.05]{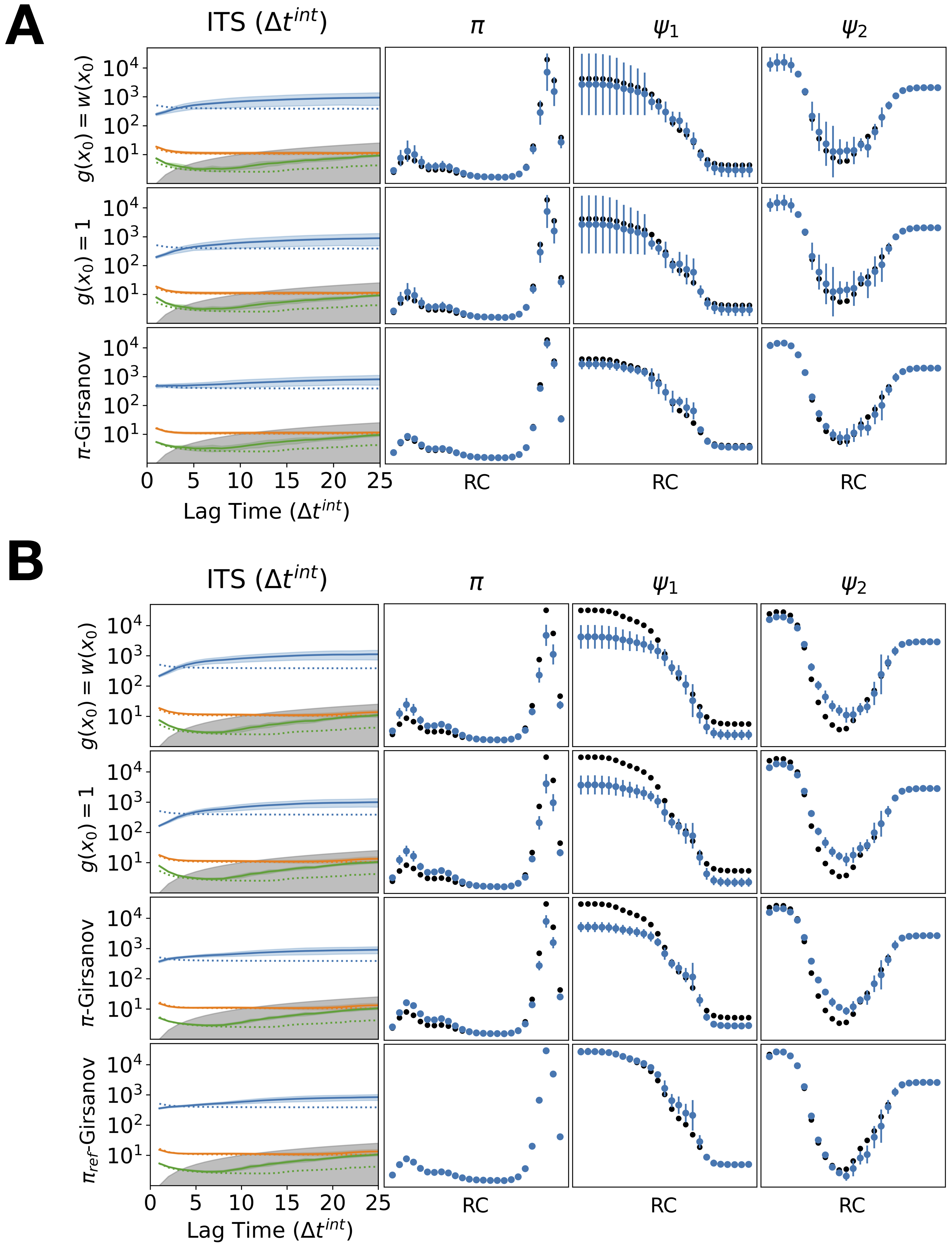}
    \caption{Summary of results for 2D M\"{u}ller-Brown potential rerun and build-up simulations: (A) The ITS, the stationary distribution $\pi$ and the leading eigenvectors $\psi_1$, $\psi_2$, $\psi_3$ of $\mathbf T(\tau)$ reweighted with different methods from the rerun metadynamics trajectories. (B) The ITS, the stationary distribution $\pi$ and the leading eigenvectors $\psi_1$, $\psi_2$, $\psi_3$ of $\mathbf T(\tau)$ reweighted with different methods from the build-up metadynamics trajectories. For all ITS plots, we use dotted lines to represent the reference values obtained from the reference MSM, solid lines to represent the values from the reweighted MSM and shade to represent the variance. For all eigenvector plots, we use black dots to represent the reference values and blue dots along with error bars to represent the reweighted values and their variance. For each column of each subplots, the y-axis scale is set to uniform so comparison can be made.}
    \label{fig:5}
\end{figure}

\begin{figure}[h]
    \centering
    \includegraphics[scale=0.05]{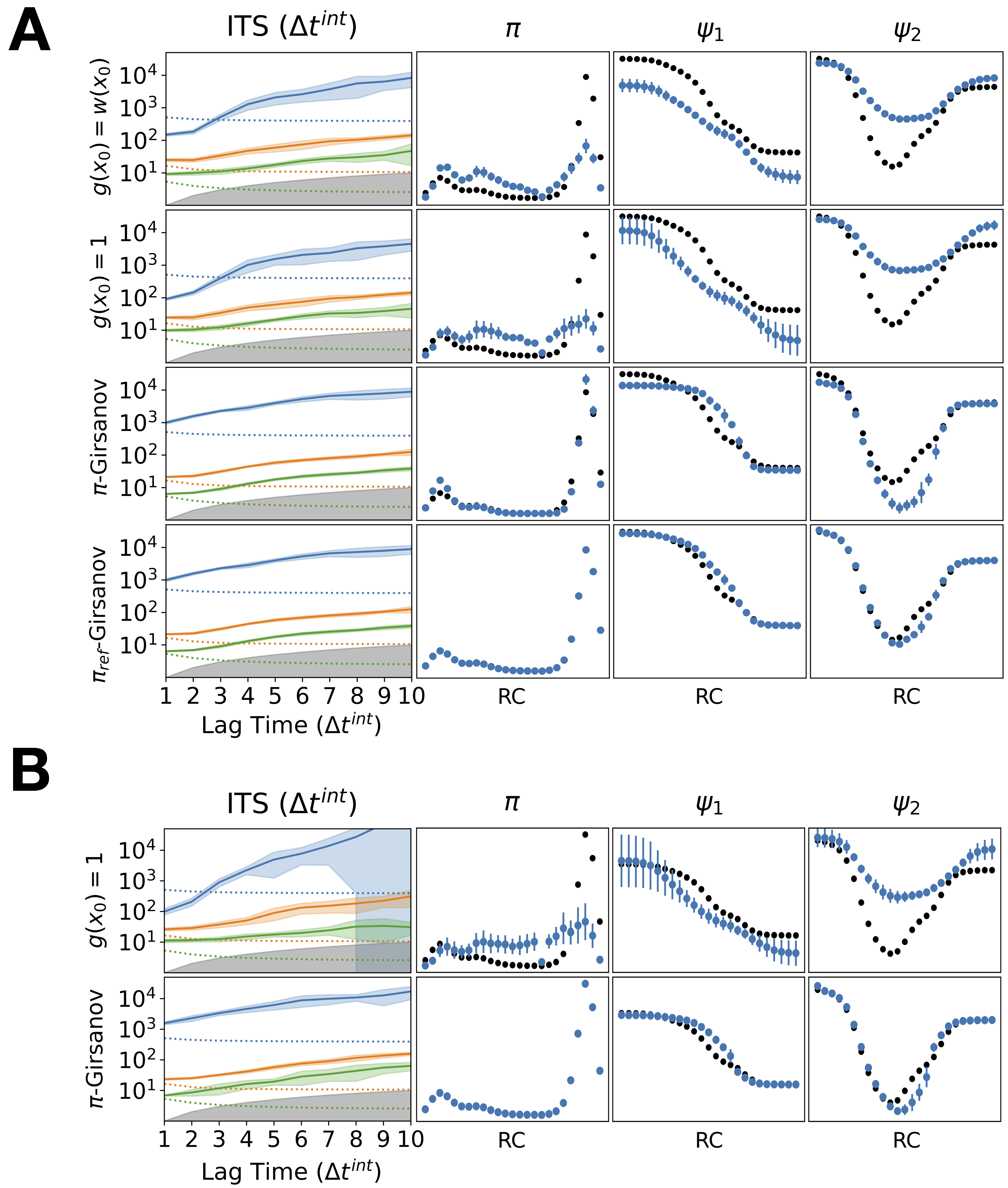}
    \caption{Summary of results for 2D M\"{u}ller-Brown potential umbrella sampling and steered MD simulations: (A) The ITS, the stationary distribution $\pi$ and the leading eigenvectors $\psi_1$, $\psi_2$, $\psi_3$ of $\mathbf T(\tau)$ reweighted with different methods from the umbrella sampling trajectories. (B) The ITS, the stationary distribution $\pi$ and the leading eigenvectors $\psi_1$, $\psi_2$, $\psi_3$ of $\mathbf T(\tau)$ reweighted with different methods from the steered MD trajectories. For all ITS plots, we use dotted lines to represent the reference values obtained from the reference MSM, solid lines to represent the values from the reweighted MSM and shade to represent the variance. For all eigenvector plots, we use black dots to represent the reference values and blue dots along with error bars to represent the reweighted values and their variance. For each column of each subplots, the y-axis scale is set to uniform so comparison can be made.}
    \label{fig:6}
\end{figure}

We further tested the $\pi$-Girsanov and the other methods on the 2D M\"{u}ller-Brown potential (Figure 4). Specifically, we performed a dimensionality reduction of the system from 2D to 1D to leave some orthogonal degree unbiased in the enhanced sampling simulations. We then use this 1D reduced coordinate as our CV $r(q)$ for all enhanced sampling simulations. This allows us to mimic the scenario of enhanced sampling simulations of realistic molecular systems, which often has most of their degrees of freedom unbiased, and to compare the capability of $\pi$-Girsanov method with other methods in such case. Since the main purpose of this paper is to test kinetic reweighting techniques, such dimensionality reduction is achieved by an ideal non-linear transformation of the original space into an optimal RC space learned from one long unbiased trajectory of the unbiased reference data set with our previously developed RC deep learning algorithm FMRC\cite{zhang2024flow}. FMRC has been previously demonstrated to perform very well as the dimensionality reduction technique before phase space discretization for MSM construction with high-dimensional trajectories of multiple complex solvated biomolecular systems\cite{zhang2024flow}. The learned RC and the projected FES onto this RC is presented in Figure 4 and we can see it optimally distinguishes the three metastable states in the system and represents the progress of transitions between these states correctly.

We then discretized only this 1D RC space to assign different coordinates of the system from different enhanced sampling trajectories into different microstates. For this system with three metastable states, we focus on the comparison of the stationary distribution $\pi$, the leading eigenvectors $\psi_1$, $\psi_2$ and their associated timescales $t_1$, $t_2$ across different reweighting and enhanced sampling methods. These results are summarized in Figure 5 and 6. For build-up metadynamics and umbrella sampling simulations, we noticed that the $\pi$ estimates from thermodynamic reweighting using (26) and (27) has not converged (Figure 5B and Figure 6A) to the reference value due to the limited amount of sampling and we included another set of MSMs built with the $\pi$-Girsanov method with a reference $\pi_{\text{ref}}$ from the unbiased simulations for our analysis to evaluate the effect of a biased, non-converged $\pi$ as constraint. We denote this method as $\pi_{\text{ref}}$-Girsanov. 

Either rerun or build-up metadynamics simulations are able to obtain decent estimates of the ITS of the leading slow modes of the system across a wide range of lag time $\tau$, whereas for umbrella sampling and steered MD simulations all three methods experience significant difficulties in both obtaining an accurate estimates of ITS and converge the $\mathbf T(\tau)$ estimation using (9) or (11) with a larger $\tau$. Overall, for this specific sample, only minor improvements in the variance of the ITS was observed for the $\pi$-Girsanov method in comparison to the original Girsanov reweighting method for rerun and build-up metadynamics simulations. The improvements in the variance of ITS are more significant for umbrella sampling and steered MD simulations, yet not very meaningful since these estimates possess a significant bias from the reference values. 

For rerun metadynamics (Figure 5A), once again, thermodynamic reweighting obtains a more accurate $\pi$ estimates using (26) and (27) than obtaining $\pi$ from the first left eigenvector of $\mathbf T(\tau)$ estimated with the original Girsanov reweighting method. As a result, the $\pi$-Girsanov method is able to obtain an unbiased estimation of $\psi_1$ and $\psi_2$ with significantly less variance, similar as above. 

In contrast, the estimation of $\psi_1$ and $\psi_2$ with $\pi$-Girsanov from the build-up metadynamics simulation is less accurate for the M\"{u}ller-Brown potential in comparison to previous examples. This is primarily due to the fact that $\pi$ estimates from (26) and (27) has not converged to the correct values and bias was introduced into the estimation of $\mathbf T(\tau)$. One can see that once an accurate $\pi_{\text{ref}}$ is used as the constraint, the deviations of $\psi_1$ and $\psi_2$ from the reference values are effectively reverted. In addition, using the same amount of simulation budget, even if the $\pi$ constraint is biased, the $\pi$-Girsanov method still outperforms the original Girsanov reweighting method, as they are almost identically deviated from the reference value but the estimates of $\psi_1$ and $\psi_2$ again have a significantly less variance.

Similar to the build-up metadynamics case, the $\pi$ estimates from the umbrella sampling trajectories of this system with MBAR has not converged. Nevertheless, one can see this estimates is still significantly better than the $\pi$ estimates obtained by the original Girsanov reweighting method, either assigning $g(x_0)$ with MBAR or ignoring $g(x_0)$. We have seen above that the attempts to obtain an accurate estimation of $t_1$ and $t_2$ from either umbrella sampling or steered MD trajectories with all three kinetic reweighting methods have failed. Similarly, the resulting $\psi_1$ and $\psi_2$ of these $\mathbf T(\tau)$ are also significantly deviated from the reference values. However, while the introduction of an accurate $\pi$ constraint was not able to improve the estimates of $t_1$ and $t_2$, it significantly improves the estimates of $\psi_1$ and $\psi_2$ and these resulting eigenvectors only deviate slightly from the reference values. This suggests the bias in $\psi_1$ and $\psi_2$ mainly arose from the bias in the $\pi$ constraint and it can be reverted with a more accurate $\pi$ constraint.

For steered MD simulations, $\psi_1$ and $\psi_2$ can also be obtained with minimal deviation if an accurate $\pi$ is provided. In comparison, if we did not apply this constraint and ignore the starting point weight $g(x_0)$, the resulting estimates of $\pi$, $\psi_1$ and $\psi_2$ are severely biased.

\subsection{Application to Alanine Dipeptide}

\begin{figure}[h]
    \centering
    \includegraphics[scale=0.05]{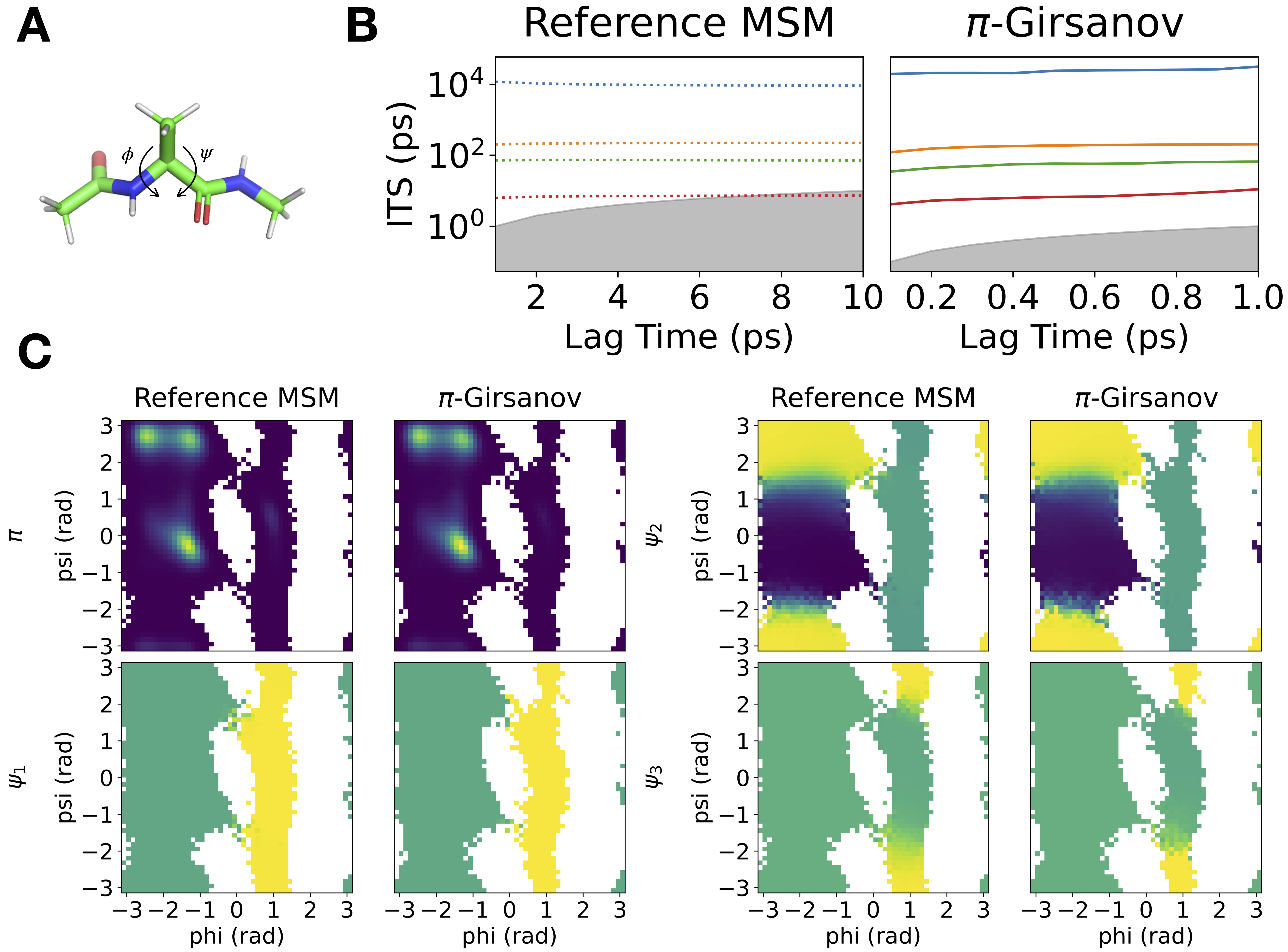}
    \caption{Summary of results of applying $\pi$-Girsanov to the alanine dipeptide system. (A) The alanine dipeptide molecule and its torsion angles $\phi$ and $\psi$. (B) The ITS of the leading eigenvectors of the reference MSM and the reweighted MSM built with $\pi$-Girsanov. The y-axis scale has been set to the same for both plots for comparison. (C) The leading eigenvectors of the reference MSM and the reweighted MSM built with $\pi$-Girsanov. Color represents the value of the eigenvectors. For each pair of plots for each eigenvector, the color scale has been set to the same for comparison.}
    \label{fig:7}
\end{figure}

We proceed to apply the $\pi$-Girsanov method to a biomolecular system: alanine dipeptide simulated in implicit solvent to demonstrate the potential of the $\pi$-Girsanov method in applications. alanine dipeptide is a well-studied small biomolecule and its dihedral angles $\phi$ and $\psi$ (Figure 7A) are known good CVs that correctly describe the system's slow mode of transitions between its metastable states.

To demonstrate the capability of our $\pi$-Girsanov method, here we specifically run two parallel build-up metadynamics simulations, each with one of the dihedral angles, $\phi$ or $\psi$, as $r(q)$ for the bias potential. This protocol is based on our results on analytical potentials above that build-up metadynamics generates reasonable estimates of $\mathbf T(\tau)$ consistently in most scenarios, while requiring no prior knowledge of the converged $b^{\text{MetaD}}_{t\to\infty}(q)$ or $\pi(r(q))$. Although in principle one can define a 2D $r(q)$ and perform a regular build-up metadynamics simulation directly, here we would like to use this setup to further demonstrate the flexibility of our $\pi$-Girsanov method on the most general setup we aimed for: the scenario where biased trajectories come from non-equilibrium simulations with different time-dependent bias potentials. 

One can see that it is very challenging to determine $g(x_0)$ under this setup. Nevertheless, as stated above in Methods, if we follow the assumption that each time-dependent bias potential of these simulations are pseudostatic, we can apply MBAR to obtain $w(x)$ and compute the stationary distribution $\pi$. We can then ignore $g(x_0)$ and apply the $\pi$-Girsanov method to these trajectories to build MSM to obtain kinetic information of the system.

The results are summarized in Figure 7. We can see that the $\pi$ estimates from combining both build-up metadynamics trajectories with MBAR is very close to the $\pi$ estimates from our reference unbiased simulations. Applying the $\pi$-Girsanov method to these trajectories, we recover the timescales $t_1$, $t_2$ and $t_3$ of the three major slow modes of the system in a good agreement with the reference values, with only a slight overestimation of $t_1$. The eigenvectors $\psi_1$, $\psi_2$ and $\psi_3$ from the reweighted $\mathbf T(\tau)$ are also in good agreement. The only limitation here is still that a large $\tau$ is not allowed for $\pi$-Girsanov due to the variance in $M(\omega)$. Yet, one can see $t_1$, $t_2$ and $t_3$ have well converged in such limited range of $\tau$. Such results also demonstrate the effectiveness of our $\pi$-Girsanov method in increasing statistical efficiency of building MSMs, since we only performed 0.1 $\mu$s biased simulations of the system in comparison to the 5 $\mu$s simulation time for the unbiased reference trajectories.

\section{Discussion and Conclusion}

In this paper, we presented the main challenge of applying the original Girsanov reweighting method to multiensemble and non-equilibrium enhanced sampling simulations and proposed a new generalized $\pi$-Girsanov method that allows the estimation of $\mathbf T(\tau)$ from enhanced sampling trajectories simulated under single or multiple different biases that are either time-dependent or independent. 

The core idea of the $\pi$-Girsanov method is to ignore the starting point weight $g(x_0)$, which can be very hard to compute for multiensemble and non-equilibrium enhanced sampling trajectories, and first compute a reweighted but biased cross-correlation matrix $\hat {\mathbf  C}(\tau)$. The bias in $\hat {\mathbf  C}(\tau)$ can then be effectively removed by a $\pi$-constrained MLE of the transition matrix $\mathbf T(\tau)$ as shown in the previous and current studies, if an accurate estimate of $\pi$ can be provided. Thanks to the well-established methods for thermodynamic reweighting, obtaining such accurate $\pi$ with trajectories from these enhanced sampling methods are often straight-forward. 

The effectiveness of the $\pi$-Girsanov method was demonstrated first on biased trajectories from multiple popular enhanced sampling simulations on analytical potentials. Our results on 1D and 2D analytical potentials suggest that the choice of enhanced sampling method which generates the trajectories has a significant impact on the quality of the final MSM constructed in terms of its ITS and its eigenvectors for both the original and our $\pi$-Girsanov method. Overall, if one already has a good estimate of $\pi(r(q))$ to construct a decent rerun metadynamics bias potential $b^{\text{MetaD}}_{t\to\infty}(q)$, then trajectories from simulations with this static $b^{\text{MetaD}}_{t\to\infty}(q)$ is preferred for both methods. However, in reality, this is often not the case and one often need preliminary e.g. build-up metadynamics simulation to obtain this $b(q)$. We show that the $\pi$-Girsanov method is able to directly construct accurate, high-quality MSMs from build-up metadynamics simulations once the $\pi$ estimate from thermodynamic reweighting is converged and no further rerun simulation is needed. 

In contrast, reweighting umbrella sampling and steered MD simulations with $\pi$-Girsanov, which uses a harmonic bias potential that localizes the trajectory to a small region of the phase space, can be very challenging, especially at large $\tau$. At the current stage, we postulate the difficulties arise from the use of harmonic bias potential with a large spring constant $\kappa$ instead of the $\pi$-Girsanov method itself. The harmonic bias potential restricts the paths to a narrow window. For lag times that are long enough for the system to move out of this window on the unbiased potential, the path ensemble sampled in umbrella sampling quickly diverges from the unbiased path ensemble. The small overlap between the umbrella sampling path ensemble and the unbiased path ensemble leads to a poor estimate of the reweighted $\mathbf T(\tau)$. This issue may also arise for rerun and build-up metadynamics simulations if we use a stronger bias (e.g. by increasing $\gamma$) as previously seen in Ref.~\citenum{donati2018girsanov}. After all, these popular enhanced sampling methods are originally designed to generate phase space samples effectively for estimating phase space ensemble average instead of path space samples for estimating path space ensemble average. Therefore, the resulting biased path ensemble may not be optimal for effectively sampling path samples, even if transitions across barriers are promoted. Furthermore, as seen in our results, the situation can be even worse if we use a large $\tau$ as the volume of the path space grows exponentially with $\tau$, especially when sampling efficiency in the path space is low. 

The final part of our results on alanine dipeptide further showcases the flexibility and the potential of the method in biomolecular applications. In this example, we demonstrate that the $\pi$-Girsanov method is able to effectively combine path samples from multiple enhanced sampling simulations of alanine dipeptide under different time-dependent biases to estimate $\mathbf T(\tau)$. The resulting $\mathbf T(\tau)$ from biased simulations is able to align with the reference $\mathbf T(\tau)$ from much longer unbiased simulations and correctly represent the system slow modes and their corresponding timescales. However, we do acknowledge that similar to what we have seen in the analytical potential cases and in previous studies\cite{donati2017girsanov,donati2018girsanov,schafer2024implementation}, estimating $\mathbf T(\tau)$ at a large $\tau$ is still very challenging due to the reasoning above.

Improving the efficiency of path ensemble sampling to further improve the applicability of the $\pi$-Girsanov method at larger $\tau$ is out of the scope of this paper. Yet, we would like to point out that there have been recent studies suggesting that one can use the idea of stochastic optimal control to perform effective enhanced sampling in the path space\cite{zhang2013importance,hartmann2018importance,holdijk2023stochastic}. Such simulations, controlled with a time-dependent bias potential, are clearly non-equilibrium simulations which does not converge to a unique stationary distribution similar to the steered MD simulations. Our results on steered MD simulations of all three systems suggest one may bypass this difficulty with $\pi$-Girsanov, once one has obtained a decent estimation of $\pi$ through standard enhanced sampling methods like umbrella sampling or metadynamics. Furthermore, while biased trajectories from these methods may not sample the path samples effectively, these path samples can still be combined to further improve the estimation of $\mathbf T(\tau)$ under the $\pi$-Girsanov framework. We leave these possible improvements for future studies. 

Taken together, we believe that the $\pi$-Girsanov method proposed here is a major step in the development of kinetic reweighting techniques for MSM construction from enhanced sampling simulations considering its improved flexibility and effectiveness in comparison to the original Girsanov reweighting method.

\appendix

\renewcommand{\theequation}{\Alph{section}\arabic{equation}}
\setcounter{equation}{0}

\section{Relationship between $\rho(x_{t+\Delta t}|x_t)$ and $\rho(\eta_t)$}
We can consider each integration cycle of the ABOBA integrator described in (\ref{eq:3}) as a function which maps the system state $x_t = (p_t,q_t)$ and the random number $\eta_t$ at time $t$ to the system state $x_{t+\Delta t} = (p_{t+\Delta t},q_{t+\Delta t})$ at time $t+\Delta t$. Then, following the derivation in the Appendix of Ref.\citenum{kieninger2023girsanov}, we can compute the Jacobian $\mathbf J$ of $x_{t+\Delta t}$ with respect to $\eta_t$ as:
\begin{equation}
\mathbf J = 
\frac{\partial x_{t+\Delta t}}{\partial \eta_t} 
= 
\left[
\begin{matrix}
\frac{\partial p_{t+\Delta t}}{\partial \eta_t} \\
\frac{\partial q_{t+\Delta t}}{\partial \eta_t}
\end{matrix}
\right]\\
= \left[
\begin{matrix}
\sqrt{k_BTM(1-e^{-2\xi\Delta t})} \\
\frac{1}{2}\Delta tM^{-1}\sqrt{k_BTM(1-e^{-2\xi\Delta t})}
\end{matrix}
\right]\\
\label{eq:a1}
\end{equation}
Applying the change-of-variables formula, we have
\begin{equation}
\begin{aligned}
\rho(x_{t+\Delta t}|x_t) 
&= \rho(\eta_t) \cdot \det\left[(\mathbf J^\top \mathbf J)^{-1/2}\right]\\
&= \rho(\eta_t) \cdot \det\left[\left(k_BTM(1-e^{-2\xi\Delta t})+\frac{1}{4}\Delta t^2 k_BTM^{-1}(1-e^{-2\xi\Delta t})\right)^{-1/2}\right]\\
&= \rho(\eta_t) \cdot C \\
& \propto \rho(\eta_t)
\end{aligned}
\label{eq:a2}
\end{equation}
where $C$ is a constant independent of $\eta_t$, $x_t$ and $x_{t+\Delta t}$. Therefore, for each integration cycle, we have
\begin{equation}
\frac{\rho(x_{t+\Delta t}|x_t)}{\rho^b(x_{t+\Delta t}|x_t)} 
= \frac{\rho(\eta_t) \cdot C }{\rho(\eta^b_t)\cdot C } = \frac{\rho(\eta_t)}{\rho(\eta^b_t)}
= \frac{\prod^{3N}_{i=1}\exp\left(-\frac{{(\eta_{t,i})}^2}{2}\right)}{\prod^{3N}_{i=1}\exp\left(-\frac{{(\eta_{t,i}^b)}^2}{2}\right)}
\label{eq:a3}
\end{equation}
which eventually leads to the formulation of $M(\omega)$ in (\ref{eq:18}).

\section{Relaxing the Equilibrium Sampling Requirement in the $\pi$-Girsanov Method}
\setcounter{equation}{0}

In this section, we show that the objective function of the proposed
$\pi$-Girsanov method remains well-defined even when the initial
configurations are not sampled from the equilibrium distribution of the
biased system. This provides a theoretical justification for applying
the method to nonequilibrium simulation data.

Let
\begin{equation}
\kappa_{\tau}(x_{0},x_{\tau})
=
\int
\mu(x_{1:\tau}\mid x_{0})
\,\mathrm{d}x_{1:\tau-1},
\end{equation}
denote the unbiased transition density, i.e., the conditional density
of $x_{\tau}$ given the initial state $x_{0}$ in the unbiased
simulation, where
$x_{1:\tau-1}=(x_{1},x_{2},\ldots,x_{\tau-1})$.
The corresponding MSM approximation can be written as
\begin{equation}
    \hat{\kappa}_{\tau}(x_{0},x_{\tau})
=
\sum_{i,j}
\rho_{ij}
\frac{\pi(x_{\tau})}
{\int_{S_j}\pi(x)\,\mathrm{d}x},
\qquad
x_{0}\in S_i,\;
x_{\tau}\in S_j.
\end{equation}

Starting from the definition of $c_{ij}$ in
\eqref{eq:16}, for an arbitrary positive weighting function
$g(x)>0$, the objective function
$\mathcal{L}=\sum_{i,j}c_{ij}\log \rho_{ij}$
maximized in \eqref{eq:7} can be rewritten as
\cite{donati2017girsanov}
\begin{eqnarray}
\mathcal{L}
&=&
\mathbb{E}_{x_{0}\sim\rho_{0}(x_{0}),
\,\omega\sim\mu^{b}(\omega)}
\!\left[
\sum_{i,j}
g(x_{0})
M(\omega)
\mathds{1}_{S_i}(x_{0})
\mathds{1}_{S_j}(x_{\tau})
\log\rho_{ij}
\right]
\nonumber\\
&=&
\iint
\rho_{0}(x_{0})
\mu^{b}(x_{1:\tau}\mid x_{0})
g(x_{0})
\frac{\mu(x_{1:\tau}\mid x_{0})}
{\mu^{b}(x_{1:\tau}\mid x_{0})}
\nonumber\\
&&\qquad\qquad
\cdot
\left(
\sum_{i,j}
\mathds{1}_{S_i}(x_{0})
\mathds{1}_{S_j}(x_{\tau})
\log\rho_{ij}
\right)
\mathrm{d}x_{0}
\mathrm{d}x_{1:\tau}
\nonumber\\
&=&
\iint
\rho_{0}(x_{0})
g(x_{0})
\mu(x_{1:\tau}\mid x_{0})
\left(
\sum_{i,j}
\mathds{1}_{S_i}(x_{0})
\mathds{1}_{S_j}(x_{\tau})
\log\rho_{ij}
\right)
\mathrm{d}x_{0}
\mathrm{d}x_{1:\tau}
\nonumber\\
&=&
\iint
\rho_{0}(x_{0})
g(x_{0})
\kappa_{\tau}(x_{0},x_{\tau})
\log
\frac{
\hat{\kappa}_{\tau}(x_{0},x_{\tau})
}{
\pi(x_{\tau})/
\int_{S_j}\pi(x)\,\mathrm{d}x
}
\,
\mathrm{d}x_{0}
\mathrm{d}x_{\tau}
\nonumber\\
&=&
-
\int
\rho_{0}(x_{0})
g(x_{0})
\,
\mathrm{KL}
\!\left(
\kappa_{\tau}(x_{0},\cdot)
\|
\hat{\kappa}_{\tau}(x_{0},\cdot)
\right)
\mathrm{d}x_{0}
\nonumber\\
&&
+
\iint
\rho_{0}(x_{0})
g(x_{0})
\kappa_{\tau}(x_{0},x_{\tau})
\log
\frac{
\kappa_{\tau}(x_{0},x_{\tau})
\int_{S_j}\pi(x)\,\mathrm{d}x
}{
\pi(x_{\tau})
}
\,
\mathrm{d}x_{0}
\mathrm{d}x_{\tau},
\label{eq:L-decomposition}
\end{eqnarray}
where
\begin{equation}
    \mathrm{KL}
\!\left(
\kappa_{\tau}(x_{0},\cdot)
\|
\hat{\kappa}_{\tau}(x_{0},\cdot)
\right)
=
\int
\kappa_{\tau}(x_{0},x_{\tau})
\log
\frac{
\kappa_{\tau}(x_{0},x_{\tau})
}{
\hat{\kappa}_{\tau}(x_{0},x_{\tau})
}
\,\mathrm{d}x_{\tau}
\end{equation}
denotes the Kullback--Leibler divergence between the true and approximate
conditional distributions of $x_{\tau}$ conditioned on $x_{0}$.

Since the second term on the right-hand side of (\ref{eq:L-decomposition}) is independent of $\rho_{ij}$,
maximizing $\mathcal{L}$ is equivalent to minimizing
$
\mathrm{KL}
\!\left(
\kappa_{\tau}(x_{0},\cdot)
\|
\hat{\kappa}_{\tau}(x_{0},\cdot)
\right)$
for every initial state $x_{0}$, weighted by
$\rho_{0}(x_{0})g(x_{0})$.

In the conventional Girsanov reweighting framework, one typically
assumes that the initial configurations are sampled from the biased
equilibrium distribution, i.e.,
$\rho_{0}(x)=\pi^{b}(x)$, and chooses
\begin{equation}
    g(x)=\frac{\pi(x)}{\pi^{b}(x)}
\end{equation}
Under this choice,
\begin{equation}
    \mathcal{L}
=
-
\int
\pi(x_{0})
\,
\mathrm{KL}
\!\left(
\kappa_{\tau}(x_{0},\cdot)
\|
\hat{\kappa}_{\tau}(x_{0},\cdot)
\right)
\mathrm{d}x_{0}
+\mathrm{const.}
\end{equation}
In contrast, the proposed $\pi$-Girsanov method simply chooses
$g(x)\equiv1$, yielding
\begin{equation}
    \mathcal{L}
=
-
\int
\rho_{0}(x_{0})
\,
\mathrm{KL}
\!\left(
\kappa_{\tau}(x_{0},\cdot)
\|
\hat{\kappa}_{\tau}(x_{0},\cdot)
\right)
\mathrm{d}x_{0}
+\mathrm{const.}
\end{equation}
Therefore, the resulting optimization objective remains meaningful even
when the initial configurations are not sampled from equilibrium.
Instead of minimizing the KL divergence averaged over the equilibrium
distribution, the objective minimizes the same quantity averaged over
the empirical initial distribution $\rho_{0}$. This observation
explains why the proposed method can be directly applied to
nonequilibrium simulation data and is consistent with the favorable
performance observed in our numerical experiments. It is worth pointing out that the optimal choice of the weighting function $g$ remains unknown, and 
a systematic investigation of this problem is left for future work.


%
%

%

\begin{acknowledgments}
M.Z. would like to thank Joana-Lysiane Sch\"afer for insightful discussions and her assistance in using the Girsanov reweighting implementation in openmmtools. Y.W. acknowledges the financial support of the National Natural Science Foundation of China (No. 32371300). B.G.K acknowledges funding  by the Volkswagen Foundation through a Momentum grant. H.W. acknowledges the financial support of the National Natural Science Foundation of China (No. 12571463).
\end{acknowledgments}

\bibliography{reference}

\end{document}